\documentclass{article}
\usepackage[utf8]{inputenc}
\usepackage{url}
\usepackage{color}
\usepackage{graphicx}
\usepackage{bm, amsmath}
\usepackage{hyperref}
\newtheorem{df}{Definition}

\title{Estimating the parameters of epidemic spread on two-layer random graphs: \\a classical and a neural network approach}
\author{Ágnes Backhausz\footnote{Corresponding author. ELTE Eötvös Loránd University, Budapest, Hungary and Alfréd Rényi Institute of Mathematics. Email: \tt{agnes.backhausz@ttk.elte.hu}}, Edit Bognár, Villő Csiszár, \\ Damján Tárkányi, András Zempléni\footnote{ELTE Eötvös Loránd University, Budapest, Hungary, Faculty of Science, Department of Probability Theory and Statistics.}} 
\date{29 February 2024}

\begin{document}
\maketitle

\section*{Abstract}
In this paper, we study the spread of a classical SIR process on a two-layer random network, where the first layer represents the households, while the second layer models the contacts outside the households by a random scale-free graph. 
We build a three-parameter graph, called polynomial model,
where the new vertices are connected to the existing ones either uniformly, or preferentially, or by forming random triangles. We examine the effect of the graph's properties on the goodness of the estimation of the infection rate $\tau$, which is the most important parameter, determining the reproduction rate of the epidemic. 

In the classical maximum likelihood approach, to estimate $\tau$ one needs to approximate the number of SI edges between households, since the graph itself is supposed to be unobservable. Our simulation study reveals that 
the estimation is poorer at the beginning of the epidemic, for larger preferential attachment parameter of the graph, and for larger $\tau$. 
We present two heuristic improvement algorithms and establish  our method to be robust to changes in average clustering of the graph model.

We also extend a graph neural network (GNN) approach for estimating contagion dynamics for our two-layered graphs.
We find that dense networks offer better training datasets.
Moreover, GNN perfomance is measured better using the $l_2$ loss function rather than cross-entropy.

\section{Introduction}

Modelling epidemic spread has been an actively studied field in the last decades, especially in the last four years, due 
to the outbreak of the SARS-CoV-2 pandemic. There are several approaches to model and predict epidemics, such as compartment models with differential equations, or various agent-based models \cite{brauer2019mathematical, britton, simon}. In the current paper, motivated by recent work on epidemic spread on networks with several layers \cite{pellis2020systematic, stegehuis2021network, turker2023multi}, we are interested in the problem of estimating the infection parameter on a random graph consisting of two layers, based on the number of susceptible and infected vertices. Although these quantities are not always easy to track precisely during an epidemic, the estimation of the actual infection parameter can be used to provide predictions, and to decide which kind of restrictions or actions are needed for mitigation.

In general, there are several, essentially different approaches to estimate the infection parameters for an epidemic spread process on a graph. For example, KhudaBukhsh et al. 
\cite{focus} compare maximum likelihood estimates with Bayesian estimates in case of a complete graph (which can be seen as a mean-field model, without any community structure or underlying graph). Britton and O'Neill \cite{britton_oneill, oneill} developed Markov chain Monte Carlo methods for the Bayesian estimate of the infection parameter; in addition, their tools also make it possible to obtain information about the connections between the vertices, even if we do not know the underlying graph (see also a survey on approximate Bayesian methods by Kypraios et al. \cite{kypraios}). On the other hand, the methods based on neural networks are also promising for various questions in statistics, both in general and for epidemic spread. Tomy et al. \cite{reconstruct} use a graph convolutional neural network to provide an estimate for the epidemic curve from a smaller sample, that is, without testing a large proportion of the population continuously. Biazzo et al.
\cite{biazzo2022bayesian} use a Bayesian generative neural network to obtain a prediction of the epidemic curve from partially observed data.

In our paper, we examine two approaches for estimating the infection parameter of the epidemic process on different graph structures. The first approach is based on the maximum likelihood estimation (MLE) of the infection parameter, while the second approach employs the machine learning model presented by Murphy, Laurence and Allard \cite{gnn}, which was designed for statistical inference about epidemic spread on various graph structures.

The characteristics of the information that is available and inferred are essential qualities of statistical methods. In our classical MLE-based technique, which gives a direct prediction of the infection parameter, we assume that the underlying graph is sampled from a given ensemble with known properties, but knowledge about the specific graph itself is not required. The inference uses only aggregated information about the states of vertices from the whole network and over the course of the epidemic until a given point in time. In contrast, the second method uses the precise structure of the graph and state of the epidemics at least locally, since it gives an estimate not for the parameter directly, rather the state of a vertex given the state and properties of its neighboring vertices and edges. However, the prediction only requires the states from the previous timestep of a discretized SIR process so the information is localized in time.

In our computer simulations we used graphs having a two-layer structure and varying clustering coefficients, which expands the cases demonstrated by Murphy et al. \cite{gnn}. We were able to obtain information about the size  requirements 
of the training dataset for adequate GNN performance and whether the different topological structure of the graphs (either in the training set or the test set) affects the results. In general, we found that the size of the training set has a much more significant effect on the quality of the estimates than the structure of the graph (see Section \ref{sec:gnn}). In order to obtain graphs with  non-negligible clustering coefficient, we used preferential attachment graphs such as the model of Holme and Kim \cite{clustered-ba} or Ostroumova et al. \cite{clustering}.

To obtain flexible random graphs which represent both the community structure (e.g.\ households, workplaces)   of real-world networks and scale-free degree distribution, we have chosen a two-layer graph, a modification of the household model with a preferential attachment graph as a second layer. In general, for modelling epidemic spread, it is quite common to use multilayer models, where different layers 
represent different types of communities \cite{fourlayer, britton, britton_oneill, multilevel}. For example, the household model introduced and analysed by Ball and coauthors \cite{ball2015seven, household, ball2009threshold, config_sir} starts with a layer for households -- this consists of disjoint, small complete graphs -- then adds a layer representing the connections between households -- this can be either a complete graph, as in the original household model, or a random graph with a more inhomogeneous structure, closer to real-world networks.  In such models there are many interesting questions about parameter sensitivity, the effect of the 
structure of the graph on the epidemic process, or about estimating the parameters, already in the two-layer case; see 
e.g.\ \cite{britton} for a recent survey on this topic. 
In the current work we examine the effect of the preferential attachment structure and the clustering coefficient (number
of triangles) on the goodness of the estimation of the infection rate in household models, and compared classical (maximum 
likelihood) and neural network methods from this point of view. The question about the clustering coefficient is raised as an open question in Section 6.2 of \cite{britton}.

\section{The model: SIR process on a household graph with tunable clustering coefficient}
\label{sec:model}

In the sequel, we focus on two-layer random graphs, where the first layer, the graph of households is deterministic, but the 
second layer, the connections between the households are chosen randomly.  In addition, for simplicity, we keep the size of the
households fixed. In this sense this model is a mean-field model: in reality, of course, the sizes are different, and we replace all household sizes with the same average value. However, our main observations about the effect of the structure of the graphs on the quality of estimation are expected to hold for the more general setup as well, as smaller differences in the degree distribution usually do not cause significant changes neither in the epidemic process nor in the properties of our estimators. 

An important point in our studies is the clustering coefficient. Loosely speaking, we are interested 
in the effect of the probability that the neighbors of a given vertex are connected to each other. Before we define the
model, we formulate the versions of the clustering coefficient which we will use. 
\begin{df} Let $G=(V, E)$ be a simple graph. \\
The  \emph{global clustering coefficient} is the ratio of three times the number of triangles to the number of pairs of adjacent edges in $G$.\\
The \emph{average local clustering coefficient} $C(G)$ is defined as follows: for each vertex $v$, let $T_G(v)$ be the number
of triangles containing $v$, and $C_G(v)$ the number of paths of length $2$ with the middle point $v$; that is, 
$\binom{N_G(v)}{2}$, where $N_G(v)$ is the number of neighbors of $v$. 
Then \[C(G)=\frac{1}{|V|}\sum_{v\in V} \frac{T_G(v)}{C_G(v)}.\]
\end{df}
I.e. for each vertex $v$, we calculate the probability that two uniformly randomly chosen neighbors of $v$ are connected to each other with an edge, and compute the average of this probability over all vertices of the graph $G$. 

When modelling real-world networks, preferential attachment models are often used, at least as approximative models.
However, the original Barab\'asi--Albert graph
\cite{ba} is not a very good choice, 
as Bollob\'as and Riordan \cite[Theorem 12]{bollobas}, proved that the 
global clustering coefficient of a Barab\'asi--Albert graph on $n$ vertices is asymptotically equal to $c (\log n)^2/n$, which
goes to $0$ as $n$ tends to infinity. 

Hence, to study flexible models with 
realistic 
 clustering coefficients, 
 we used the 
{\it polynomial graph models} defined by Ostroumova, Ryabchenko and Samosvat \cite{clustering}. In particular, we study the 
three-parameter model described in Section 5.1 of the above paper, or its minor modifications. We start with an arbitrary graph
with $n_0$ vertices and $mn_0$ 
edges. Then, we let the graph to grow as follows:
in the $(n+1)$st step, we add a new vertex
labelled $(n+1)$ to the graph, together with $m$ 
edges, which connect the new vertex to 
the set $\{ 1, \ldots, n\}$. 
We denote the graph after $n$ steps by $G_n$.
In this particular model we choose an even number $m=2q$, and the $m$ edges are chosen in $q$ pairs, where the choice of these 
$q$ pairs is independent and identically distributed. 
The choice of one pair of edges goes as follows (we give here the notations of \cite{clustering} as well, but in the sequel
we use our own notation): 
\begin{itemize}
    \item With probability $p_{\rm pa} = \alpha$, we choose two vertices independently from $\{ 1, \ldots, n\}$, with 
    probability proportional to their in-degree (preferential attachment component), and connect a new edge to these vertices.
    \item With probability $p_{\rm u} = \delta$, we choose two vertices independently from $\{ 1, \ldots, n+1\}$, with equal
    probability (uniform component), and connect a new edge to these vertices.
    \item With probability $p_{\rm tr} = \beta$, we choose a random edge of $G_n$ uniformly, and connect a new edge to both of
    its endpoints (triangular component).
\end{itemize}
Naturally, we have $p_{\rm pa} + p_{\rm tr} + p_{\rm u} =1$. 
The authors of  \cite{clustering} also show 
that if $2p_{\rm pa} + p_{\rm tr} < 1$, then the global clustering coefficient of the graph does not tend to zero as 
$n\to\infty$, but is asymptotically
$$ \frac{3(1-2p_{\rm pa} - p_{\rm tr})p_{\rm tr}}{5m - 1 - 2(2m-1)(2p_{\rm pa} + p_{\rm tr})}.$$
In addition, the asymptotic degree distribution is scale-free with exponent $\gamma = 1 + \frac{2}{2p_{\rm pa}+p_{\rm tr}}$. 
It seems that there is no closed formula for the limit of the local average clustering coefficient in this model, but computer 
simulations show that we can tune this quantity nicely (as presented in \cite{clustering}, and in the next section we also
analyze this in more details).

We use the following minor modification of the algorithm, which also works when $m=2q+1$: first we randomize the number
of triangles, assuming a binomial distribution of order $q$ with parameter $p_{\rm tr}$, and the remaining edges are formed 
one-by-one according to preferential attachment (with probability $p_{\rm pa}/(p_{\rm pa}+p_{\rm u})$) or uniform attachment 
(with probability $p_{\rm u}/(p_{\rm pa}+p_{\rm u})$). 
Multiple edges are allowed, but when 
calculating the clustering coefficient, we take the multiplicity of each edge as $1$. 

Putting this together with the household model, we obtain the following model of epidemic spread on a two-layer random graph:
\begin{enumerate}
\item We have $N$ vertices, who live in households of size $N_{hh}$.  
In this paper we use $N_{hh}=5$. Each household is a 
complete graph with edges of weight $1$. 
\item Independently of the household layer, we create a polynomial graph model on the $N$ vertices, defined above. 
This represents the connections between households. The weight of these edges will be fixed, $w$, which is considered as a 
given constant and can be used in the estimations (in the simulations we will typically use $w=0.4$). 
\item  We run  an SIR (susceptible-infected-recovered) process on this random graph. The infection rate within households is 
$\tau$, while the infection rate on each edge connecting individuals from different households is $\tau\cdot w$. 
Our goal is to
estimate the infection rate $\tau$ given the number of S, I, R vertices and maybe some additional information.
\end{enumerate}

An alternative for the polynomial model can be the model of Holme and Kim \cite{clustered-ba}. For this model, which also has a
preferential attachment component, it is proved that the global clustering coefficient tends to zero as the number of vertices
tends to infinity, but the average local clustering coefficient has a positive limit. 

\section{Epidemic spread on graphs with different structure}

In this section we study the effect of the different parameters on the structure of the random graph and on the epidemic curve in the model defined in 
Section \ref{sec:model}, in order to find parameter setups that correspond to realistic scenarios. 

\subsection{Average local clustering coefficient}

\begin{figure}
    \begin{minipage}{0.5\textwidth}
    \centering
    \includegraphics[width=0.8\textwidth]{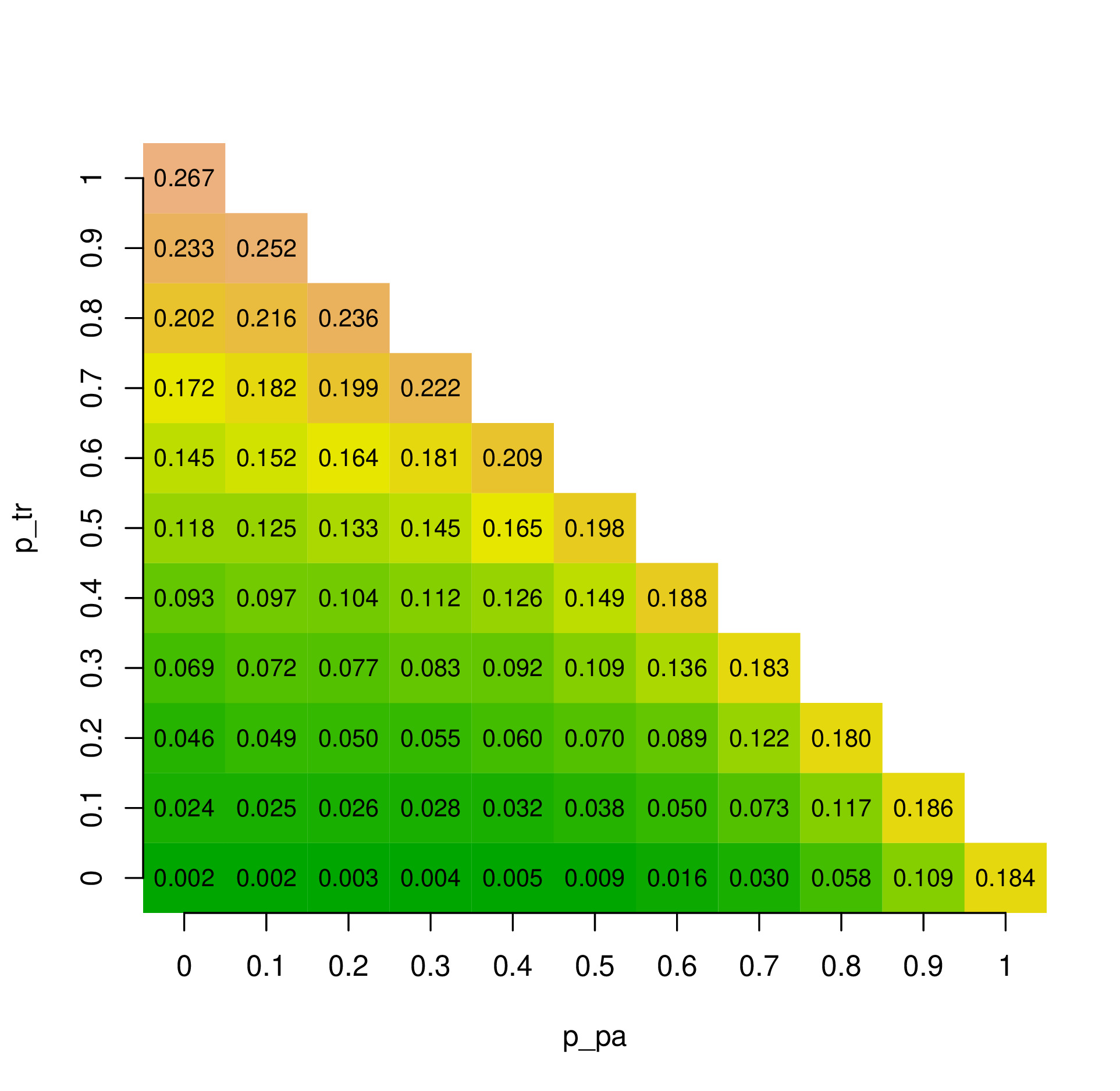}
    \end{minipage}
\begin{minipage}{0.5\textwidth}
    \centering
    \includegraphics[width=0.8\textwidth]{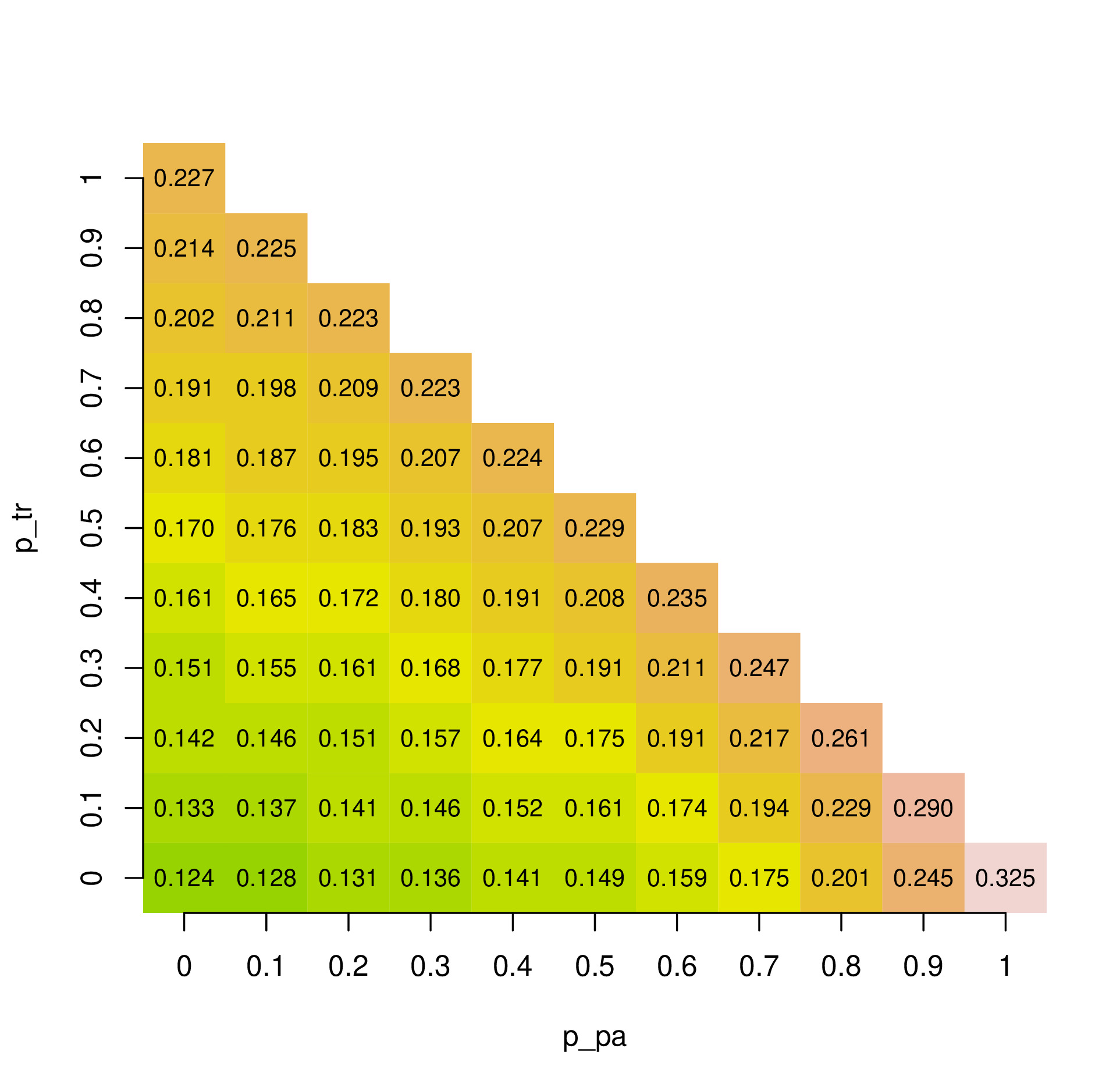}
    \end{minipage}
\caption{Average local clustering coefficient for the polynomial graph (left) and the two-layer graph (right). The rows show the value of $p_{\rm tr}$, the columns show the values of $p_{\rm pa}$. $N=5000$, $m=4$.}
\label{fig:klasztlok4}
\end{figure}

First we focus only on the graph, in particular, we study the average local clustering coefficients of the polynomial graph and two-layer graphs with the following setting: we have $N=5000$ vertices altogether, $n_0 = 50$ vertices in the initial configuration, and the average degree in the polynomial graph is approximately $8$ as $m=4$.   We were interested in the effect of the parameters controlling the probability of preferential attachment choice and creating triangles, so we changed the values of $p_{\rm pa}$ and $p_{\rm tr}$ using an equispaced grid, with $p_{\rm pa}+p_{\rm tr}\leq 1$.  In 
Figure \ref{fig:klasztlok4}, each value is the average of the local clustering coefficients for $10$ independent simulations. We can see that the average local 
clustering coefficient of the polynomial graph ranges from almost $0$ to around $0.27$ (the latter occurs when we add many triangles due to $p_{\rm tr}=1$).
 We can see that for the polynomial model, the average local clustering coefficient depends on both the triangle and the preferential component in a monotonic increasing way, and of the two, the triangle component has a stronger effect. The reason for this is that in the polynomial graph, as the new vertex is connected to old vertices, there is a small chance to obtain any triangles, unless the triangle component is high.

When we add the layer of households, i.e. small complete graphs with size $N_{\rm hh}=5$, the clustering coefficient naturally increases:
it ranges between $0.12$ and $0.33$. Observe that the average local clustering coefficient still depends on both the triangle and the preferential component in a monotonic increasing way. However, interestingly, now the preferential attachment component has a stronger effect -- at least when the uniform component is very small -- the highest clustering occurs 
when $p_{\rm pa}$ takes its maximal value $1$. 
An explanation for this could be the following. When we have a purely preferential graph, vertices with very high degree (so-called hubs) emerge. When we form the households, the neighbors of a hub get connected with each other, thus forming many triangles. 

From Figure \ref{fig:klasztlok4} we can also see that in most cases the clustering coefficient in our two-layer graphs is similar to real-world observations. Bokányi et al. \cite{bokanyi2023anatomy} analyzed a multi-layer network with registered connections of family, household, work, school and next-door neighbors. They found that the clustering coefficient is close to $0.4$. In addition,  the average local clustering coefficient of a typical online social network is between $0.1$ and $0.4$ 
\cite{closure}, which also matches our model. 

\subsection{Choice of the infection rates}

Our next goal is to compare epidemic spread on the two-layer graph with different values of $p_{\rm pa}$ and 
$p_{\rm tr}$, with fixed infection parameter $\tau$. In order to choose realistic values of $\tau$, we give a heuristic estimate of the value of $R_0$, the basic reproduction number
(the expected number of infections caused by an infected individual at the beginning of the epidemic). In one of the most intense periods of the coronavirus pandemic, in Spring
2020 in Italy, $R_0$ was estimated to be between $2.43$ and $3.1$  \cite{italy}. Hence we 
choose such $\tau$ values that $R_0$ is
between $1$ and $3$. 

Now let us see how $R_0$ depends on $\tau$. We  assume that vertex $v$ gets infected, and  
calculate $R_0$, the expected number of infections it causes. 
From \cite{tran} (see Proposition $2.2$) we know that  $R_0=(d-1)\cdot\frac{\tau}{\tau+\gamma}$ for a random $d$-regular graph (where $\tau$ is the infection rate and $\gamma$ is the recovery rate), where the subtraction corresponds to the individual that infected $v$. In our model, the  expected number of infected neighbors is different if the vertex that infected $v$ is in the same household, or if it is from a different household. We estimate the probability that a vertex was infected by a household member by the proportion of weighted degrees, which is  $(N_{\rm hh}-1)/(N_{\rm hh}-1+w\cdot 2m)$. Hence, by taking into consideration that the infection rate between households is $w\cdot \tau$, we obtain that
\begin{align*}R_0&\approx \frac{N_{\rm hh}-1}{N_{\rm hh}-1+w\cdot 2m}\cdot \bigg(\frac{(N_{\rm hh}-2)\cdot \tau}{\tau+\gamma}+\frac{2m\cdot w\cdot \tau}{w\cdot\tau+\gamma}\bigg)\\&\quad +\frac{w\cdot 2m}{N_{\rm hh}-1+w\cdot 2m}\cdot \bigg(\frac{(N_{\rm hh}-1)\cdot \tau}{\tau+\gamma}+\frac{(2m-1)\cdot w\cdot \tau}{w\cdot\tau+\gamma}\bigg)\end{align*}

Using this formula with $N_{\rm hh}=5$, $m=4$, $w=0.4$ and recovery rate $\gamma=1$, for $\tau=0.3$ we get $R_0\approx 1.6$, while for $\tau=0.6$, we get $R_0\approx 2.75$. As we have seen above,  these values are in the interval that is observed from real data.

\subsection{Effect of the preferential attachment component}

First we are interested in the effect of the preferential attachment component on the proportion of infected vertices. Later on, when we estimate the infection parameter, it will be important to see what kind of structural properties of the graphs should we take into account, and what is the necessary information to collect to get good estimates. To start with, in the two-layer model defined in Section \ref{sec:model}, we fix the weight of the triangle component ($p_{\rm tr}=0.1$), and increase the weight of the preferential attachment component: $p_{\rm pa}=0.1, 0.2, 0.3, 0.4, 0.5$. As before, we randomize two-layer graphs with $N=5000$ vertices, we fix household size $N_{\rm hh}=5$, the average degree is approximately $8$ as we choose $m=4$, and the recovery rate is $\gamma=1$. The infection rate will be $\tau=0.3$ and $\tau=0.6$: the calculation of $R_0$ showed that this leads to realistic scenarios. The set of infected individuals in the beginning is a randomly chosen $5\%$ of the whole population (chosen independently for each simulation). For each parameter setup, we randomized five graphs, and ran the SIR process five times on each of them. 

\begin{figure}
    \begin{minipage}{0.5\textwidth}
    \centering
    \includegraphics[scale=0.2]{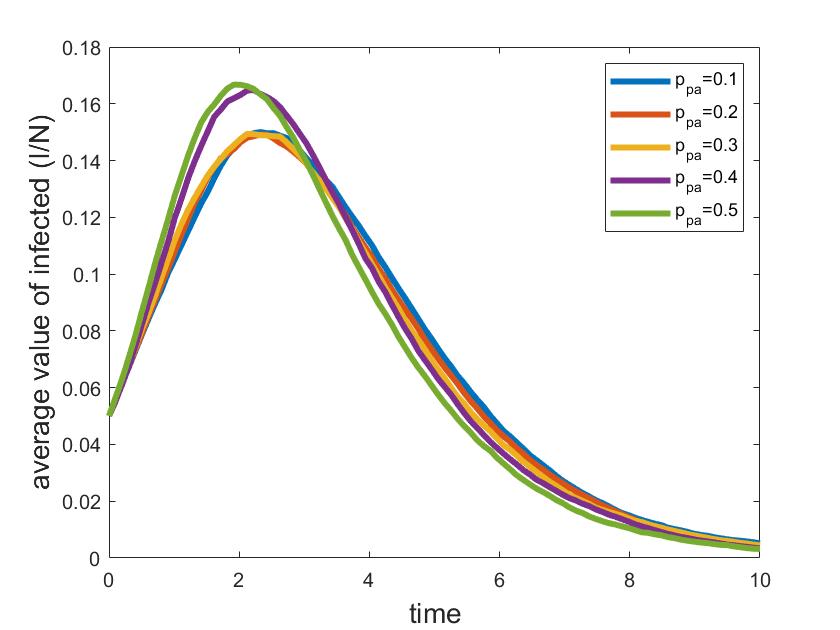}
    \end{minipage}
\begin{minipage}{0.5\textwidth}
    \centering
    \includegraphics[scale=0.2]{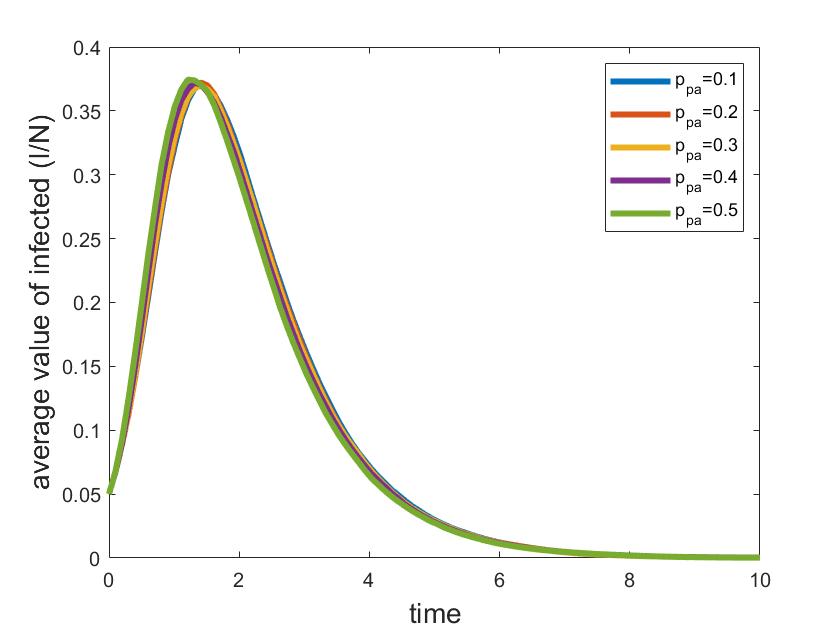}
    \end{minipage}
\caption{Proportion of infected vertices   with $p_{\rm tr}=0.1$ fixed, $N=5000$ vertices, $m=4$, $\tau=0.3$ (left) and $\tau=0.6$ (right). Each curve is the average of $25$ simulations.}
\label{fig:sensitivity-pa}
\end{figure}

Figure \ref{fig:sensitivity-pa} shows the average of the proportion of infected vertices for these 25 simulations. The preferential attachment component has a significant effect on the process for $\tau=0.3$ (at least if the weight of the preferential attachment component is at least $0.4$):
the larger the preferential attachment component is, the larger the peak of the epidemics is. This emphasises the role of vertices with large degree in the spreading process: preferential attachment dynamics lead to huge differences in the number of neighbors, and vertices with large degree have a high chance to get infected at the beginning and transmit the disease to many other vertices. However, for a more intense epidemic spread, for $\tau=0.6$, the structure of the graph does not seem to change the epidemic curve. As for the local clustering coefficient, according to Figure \ref{fig:klasztlok4}, it is essentially the same for all graphs: it is between $0.14$ (for $p_{\rm pa}=0.1$) and $0.16$ (for $p_{\rm pa}=0.5$). Hence in this setting the differences that we observe indeed come from the preferential attachment dynamics.

\subsection{Effect of the triangles}

\begin{figure}
    \begin{minipage}{0.5\textwidth}
    \centering
    \includegraphics[scale=0.2]{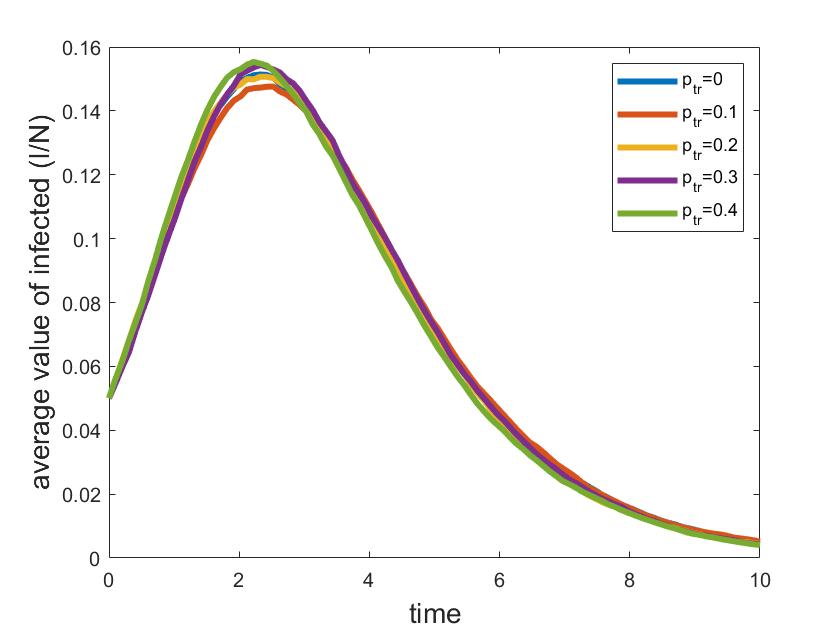}
    \end{minipage}
\begin{minipage}{0.5\textwidth}
    \centering
    \includegraphics[scale=0.2]{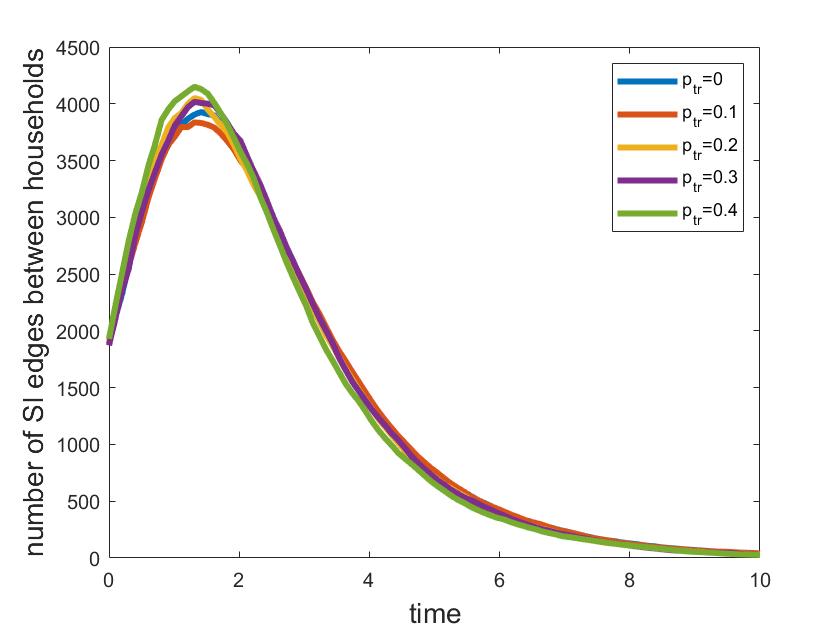}
    \end{minipage}
\caption{Proportion of infected vertices (left) and number of SI edges (right)  with $p_{\rm pa}=0.2$ fixed, $5000$ vertices, $m=4$, $\tau=0.3$. Each curve is the average of $25$ simulations.}
\label{fig:sens-tr-vertices}
\end{figure}

In addition to the vertices with large degree, the number of triangles can also have an effect on the spread of the epidemic: if the neighbors of an infected vertex are often connected to each other, they have less chance to transmit the disease to many new individuals. In our model the number of triangles is mainly governed by $p_{\rm tr}$. Hence, on Figure \ref{fig:sens-tr-vertices}, we can see the results of simulations very similar to the previous ones, but now the weight of the preferential attachment component $p_{\rm pa}=0.2$ was fixed, and we changed the weight of the triangle component from $p_{\rm tr}=0$  to $0.4$. Since it will play a crucial role in the estimates, we also included a plot of the number of edges with one $S$ and one $I$ endpoints (these are the edges along which infection is possible). In this case the estimated average local clustering coefficient goes from $0.133$ to $0.174$, according to Figure \ref{fig:klasztlok4}), so the differences are larger than in the previous case. However, there is no significant difference between the curves, which suggests that the local clustering coefficient does not have a significant impact on the spread of the epidemic.  This gives a partial answer to the open questions raised in Section 6.2 of \cite{britton}: in our setup, the clustering coefficient does not have a significant impact on the performance of the estimates.

\section{Parameter estimation based on the maximum likelihood method}

In this section our goal is to give estimates on the infection parameter, by using classical methods relying mostly on the average degree, the weight $w$, and the number of susceptible and infected individuals at each time step. We will start from a very simple approach and make some adjustments of these methods in order to study the quality of the estimates for graphs with different edge density, clustering coefficient and preferential attachment mechanism.

Our starting point is the argument of \cite{focus}, where the maximum likelihood estimation for the recovery and infection rates is determined in the case of the complete graph, which is based only on local properties of the graph. This is based only on interactions via edges, hence it is appropriate for our graph with a different structure as well. In particular, the estimate of the recovery rate is as follows:
\[\hat\gamma=\frac{z_R}{\int_{0}^{T} I_t dt}=\frac{z_R}{\sum_{t_i\leq T} I_{t_{i}} \cdot (t_i- t_{i-1})},\]
where $T$ is the time point at which we calculate the estimate; $z_R$ is the total number of events when a vertex recovers, that is, the number of recovered vertices at $T$; $S_t, I_t, R_t$ denote the number of vertices in the states $S, I, R$ at time $t$; $0<t_1<t_2<\ldots<t_i$ are the time points where there is an infection or a recovery. 

The estimate of the infection rate according to \cite{focus} is given by the following formula:
    \[\hat \tau=\frac{z_I}{\int_0^T E_t^{SI}\, dt}=\frac{z_I}{\sum_{t_i<T} E_{t_i}^{SI}(t_i-t_{i-1})},\]
    where $z_I$ is the total number of events when a vertex gets infected; $E_t^{SI}$ is the number of SI edges (edges with one susceptible and one infected endpoint) at time $t$.

\begin{figure}
\begin{minipage}{0.5\textwidth}
    \centering
    \includegraphics[scale=0.2]{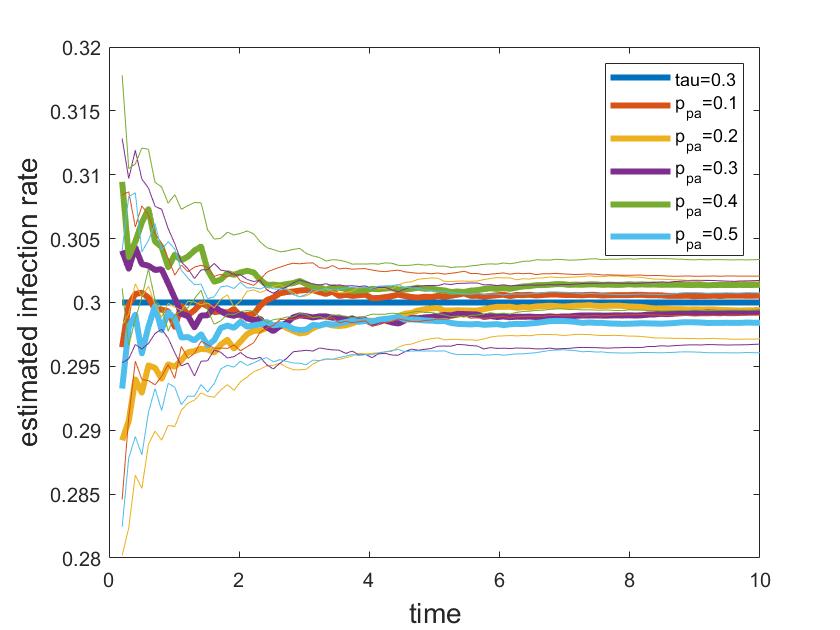}
    \end{minipage}
    \begin{minipage}{0.5\textwidth}
    \includegraphics[scale=0.2]{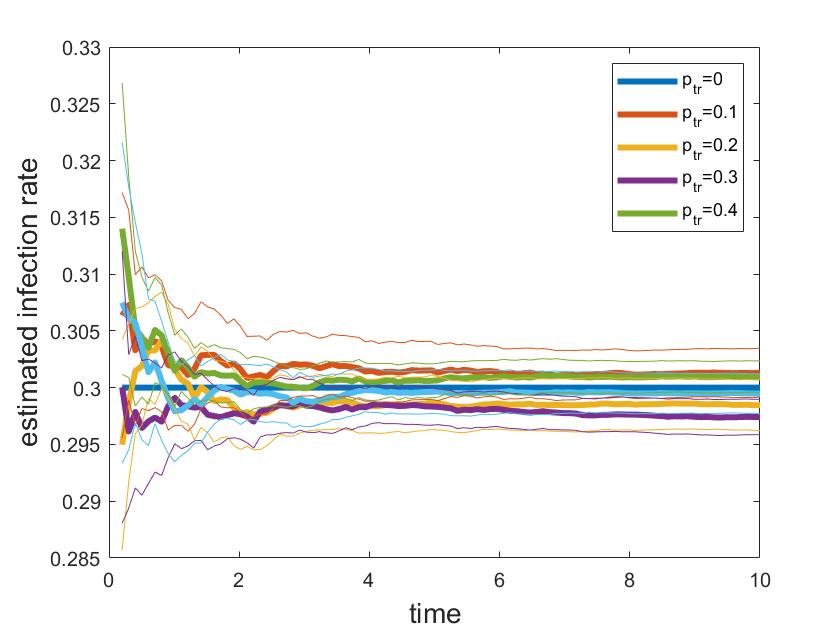}
    \end{minipage}
\caption{Estimate of the infection parameter $\tau$ for different polynomial random graphs  if the number of $SI$ edges is known exactly. The graph has $N=5000$ vertices,  $m=4$,  $\tau=0.3$, $\gamma=1$, we randomize five graphs  for each parameter setup, and run five SIR simulations on each graph; the  
thick curve is the mean of 25 trajectories, 
thin curves correspond to the $95\%$ confidence intervals based on the 25 simulations. Left: $p_{\rm tr}=0.1$ is fixed; right:  $p_{\rm pa}=0.2$ is fixed. 
}
    \label{fig:tau-pa-knownsi}
\end{figure}

Since in our model, edges have different weigths, the above formula is modified as follows: \begin{equation}\label{eq:hattau}\hat \tau=\frac{z_I}{\int_0^T W_t^{SI}\, dt}=\frac{z_I}{\sum_{t_i<T} W_{t_i}^{SI}(t_i-t_{i-1})},\end{equation}
    where $W_t^{SI}$ is the total weight  of SI edges, including the edges within households with weight $1$ and the edges between households with weight $w$.

As we can see in Figure \ref{fig:tau-pa-knownsi}, the estimated infection rate converges quickly to the real value of $\tau$, independently of the parameters of the two-layer graph model. However, this formula heavily relies on the number of SI edges, which might not be easy to measure during an epidemic, as detailed information on the graph is necessary for this.

Hence we will use the following estimate for the number of SI edges going between different households, which is based only on the number of susceptible and infected individuals:
\begin{equation} \label{si_estimate}
\hat E^{SI,o}_t=I_t\cdot \bigg(d-\frac{wd}{wd+N_{\rm hh}-1}\bigg)\cdot \frac{S_t}{N},
\end{equation}
where $d$ is the average number of the neighbors of a vertex outside its household, and $w$ is the weight of the edges going between different households (as $d$ is close to $2m$, both these quantities were considered to be known). The idea behind this formula is that we take into account each $d$ neighbors of the infected vertex with probability $S_t/N$ outside its household, but since one of these vertices could be the one that transmitted the disease to this particular vertex, we subtract the estimated probability of this event. Then the estimate of $W_t^{SI}$ is simply the number of the $SI$ edges within household plus $w$ times the  $\hat E^{SI,o}_t$, and we can use equation \eqref{eq:hattau} to get the estimate of $\tau$.

\begin{figure}
\begin{minipage}{0.5\textwidth}
    \centering
    \includegraphics[scale=0.2]{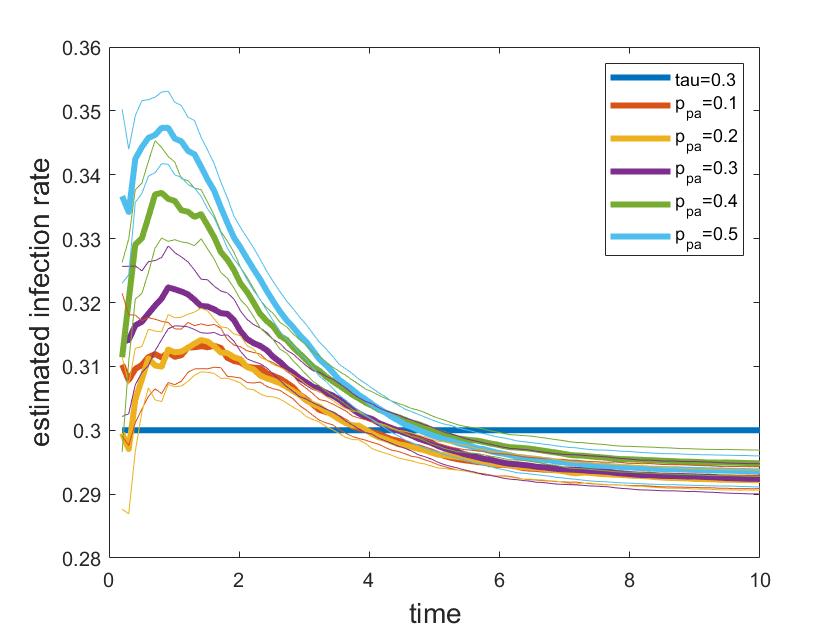}
    \end{minipage}
\begin{minipage}{0.5\textwidth}
    \centering
    \includegraphics[scale=0.2]{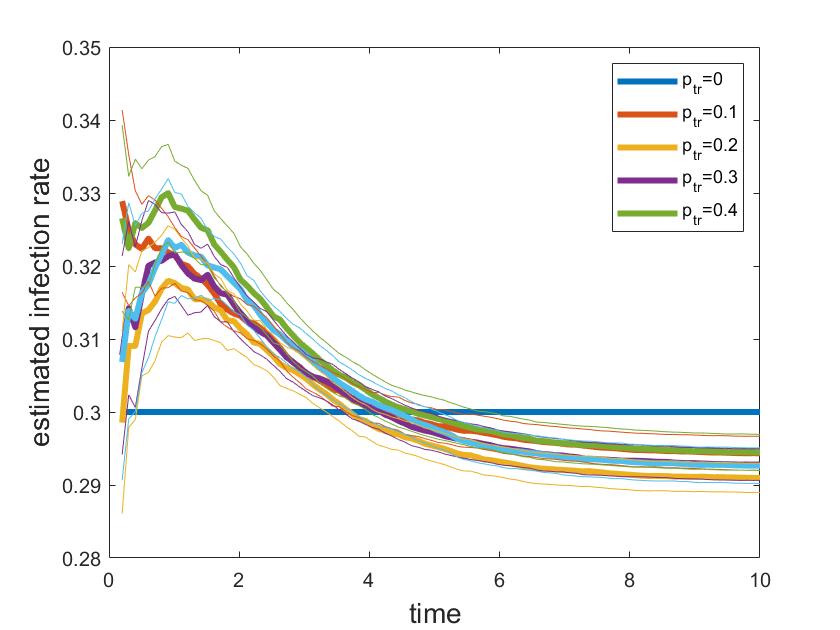}
    \end{minipage}
\caption{Estimate of the infection parameter $\tau$ for different polynomial random graphs if the number of $SI$ edges between households is not known. The graph has $N=5000$ vertices,  $m=4$,  $\tau=0.3$, $\gamma=1$, we randomize five graphs  for each parameter setup, and run five SIR simulations; the 
thick curve is the mean of 25 trajectories, 
thin curves correspond to the $95\%$ confidence intervals based on the 25 simulations. Left: $p_{\rm tr}=0.1$ is fixed; right:  $p_{\rm pa}=0.2$ is fixed. }
    \label{fig:tau-convergence-pa}
\end{figure}

On Figure \ref{fig:tau-convergence-pa}
 we can see how the formula with the estimated number of $SI$ edges works as we proceed in time. The simulation was run on graphs of $5000$ vertices, with the polynomial model with $m=4$, with $\tau=0.3$ fixed (recall that $R_0$ is approximately $1.6$ with this parameter setup). Each 
 thick curve is the average of $25$ curves, corresponding to 5 graphs for each $p_{\rm pa}$ (left) and $p_{\rm tr}$ (right) and 5 SIR simulations on each graph. The 
 thin curves represent the $95\%$ confidence intervals based on these $25$ runs. On the horizontal axis, time is measured in units (the recovery rate $\gamma=1$ is fixed). Recall from Figure \ref{fig:sensitivity-pa} that, with the same parameter setup, the peak of the epidemic is around 2. At this time, in the case when we change $p_{\rm pa}$ (left figure), the estimate is between $0.31$ and $0.35$, which is significantly worse than in the case where the total number of SI edges are known (Figure \ref{fig:tau-pa-knownsi}). We can also see that the estimate is a monotone function of the weight of the preferential attachment component: the higher the weight of the preferential component is, the higher the error of the estimate is, at least in the first part of the epidemic. Hence it is worth looking for estimates that improve \eqref{eq:hattau} by using the degree distribution or other structural properties of the graph that are closely related to the preferential attachment dynamics. On the right-hand side figure, where the weight of the preferential attachment component is fixed and the weight of the triangle component is changing, the convergence is much faster, and it does not seem to depend on the parameters. This corresponds to the fact that in this case the spread of the epidemic was also almost the same for the different values or $p_{\rm tr}$ (recall Figure \ref{fig:sens-tr-vertices}). 
 
\begin{figure}
\begin{minipage}{0.5\textwidth}
    \centering
    \includegraphics[scale=0.2]{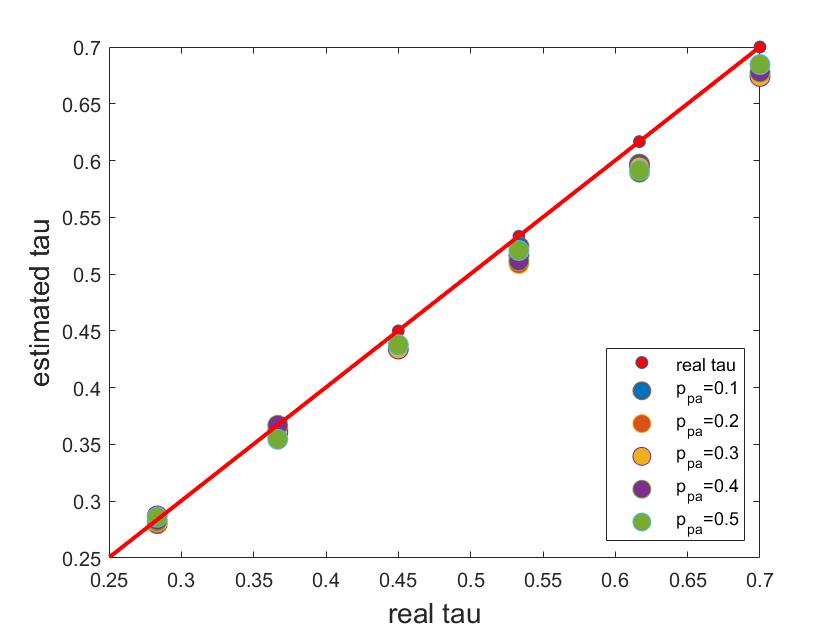}
    \end{minipage}
\begin{minipage}{0.5\textwidth}
    \centering
    \includegraphics[scale=0.2]{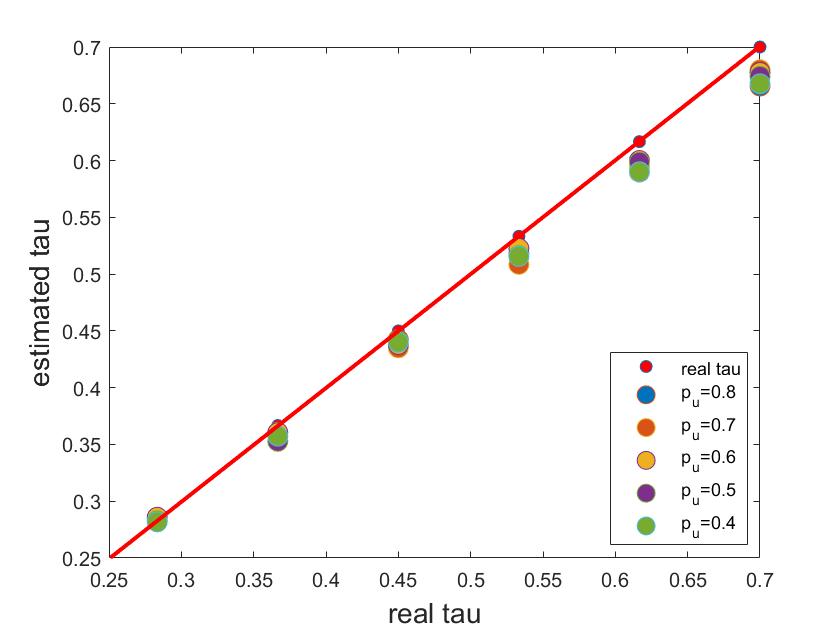}
    \end{minipage}
\caption{Estimate of the infection parameter $\tau$ for different polynomial random graphs  if the number of $SI$ between households is not known. The graph has $N=5000$ vertices,  $m=4$,   $\gamma=1$; we run five SIR simulations, the dots are averages of the five estimates when $5/6$ of the epidemic events occured. Left: $p_{\rm tr}=0.1$ is fixed; right:  $p_{\rm pa}=0.2$ is fixed. }
    \label{fig:tau-estimate-pa}
\end{figure}

Figure \ref{fig:tau-estimate-pa} shows the estimate of $\tau$ at a time which is close to the end of the epidemic: here we used $T$ such that $5/6$ of the total epidemic events (infection or recovery) occur before $T$. 
We can see that the estimate works quite well for different values of $\tau$, independently of the parameters of the graph, but it gets worse if we increase the value of the infection parameter. In the next sections we will present methods that lead to better results than this simpler method.

\subsection{A time-shifted estimate of the number of SI edges} 

In this section, we consider another estimation of the between-household SI edges, which can be used instead of equation \ref{si_estimate}.

\begin{figure}
\begin{minipage}{0.5\textwidth}
    \centering
    \includegraphics[width=0.8\textwidth]{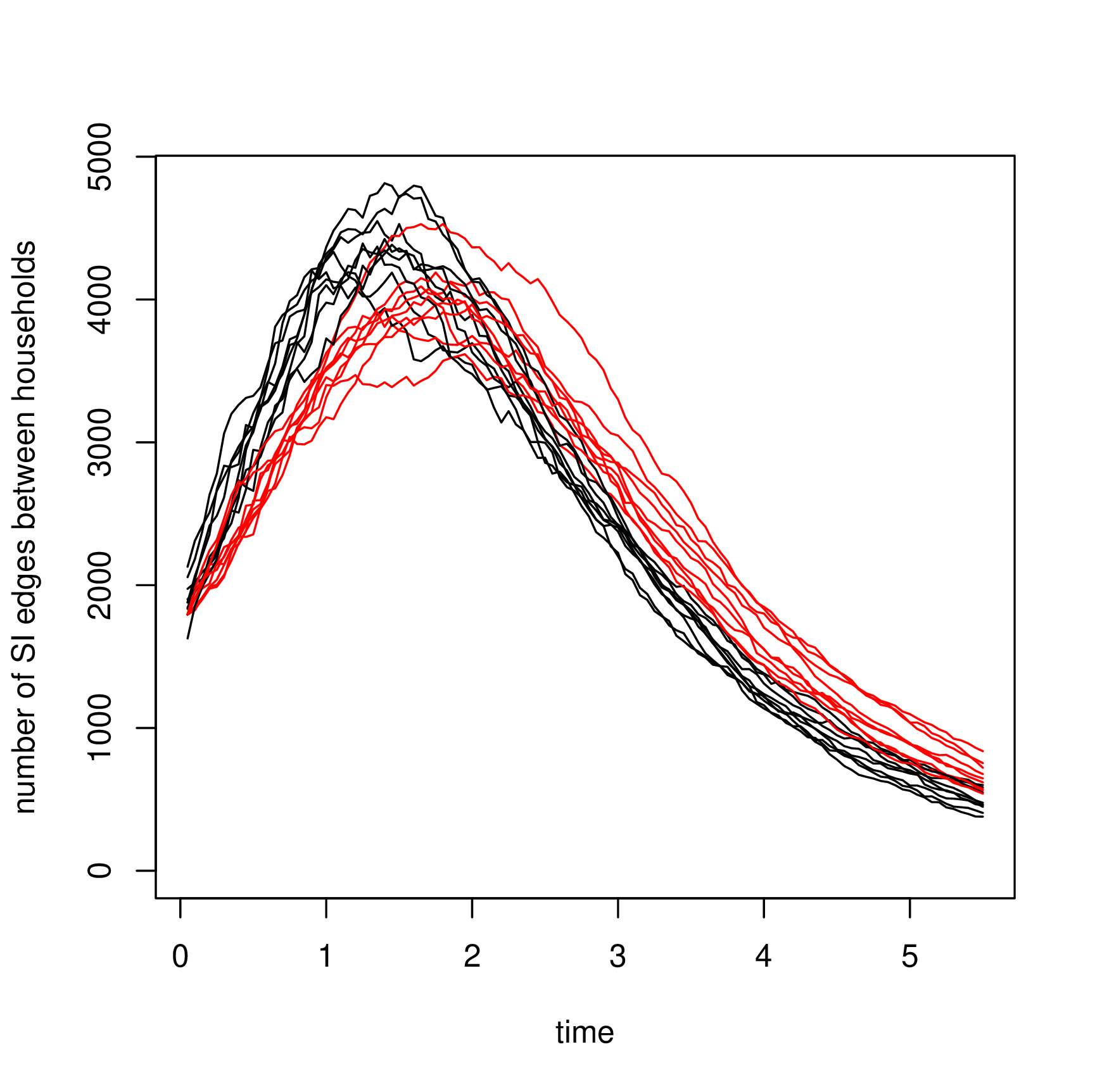}
    \end{minipage}
\begin{minipage}{0.5\textwidth}
    \centering
    \includegraphics[width=0.8\textwidth]{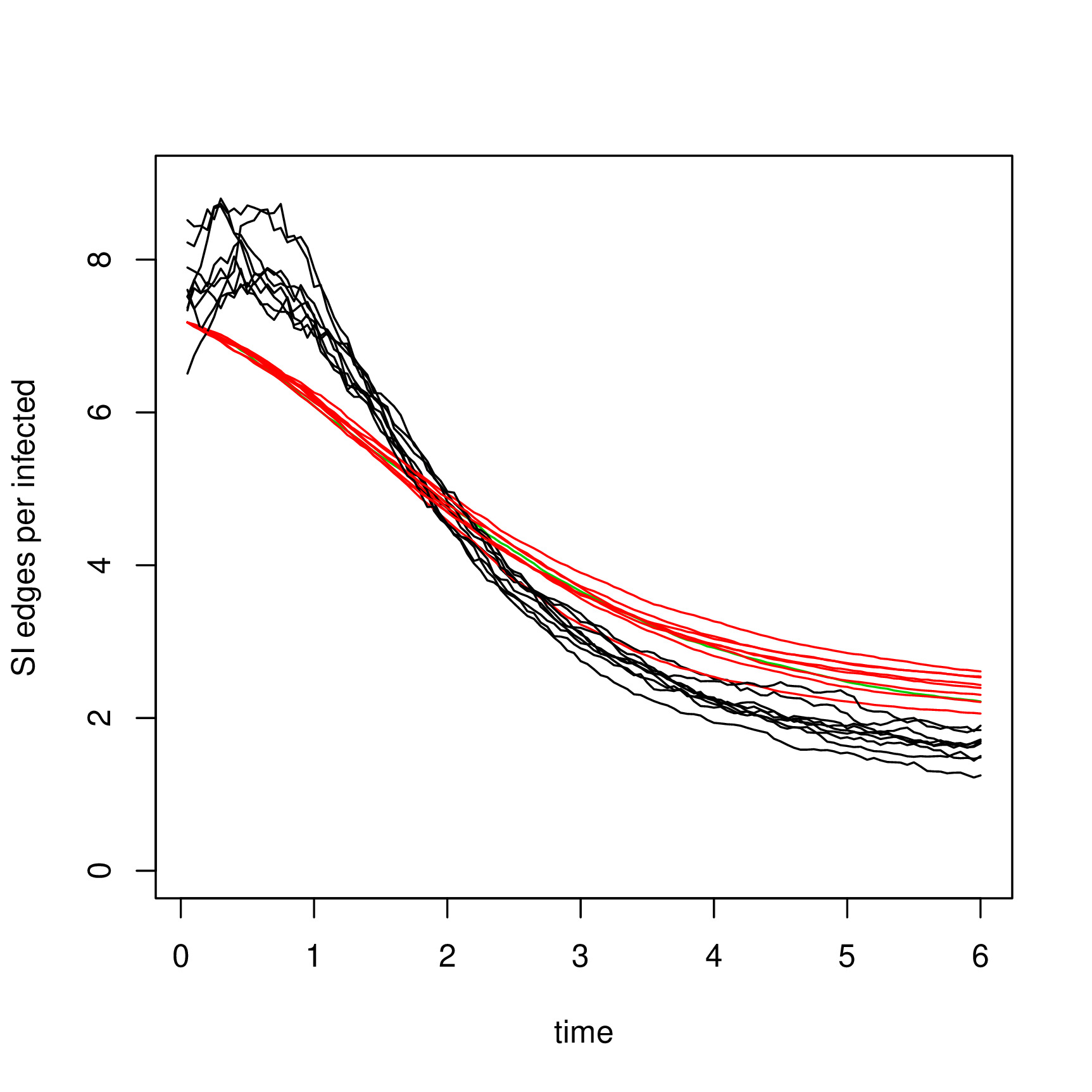}
    \end{minipage}
 \caption{Left: Number of SI edges between households (black) and their estimated number (red). 
 Right: Number of between-household SI edges per infected vertex (black) and their estimated number (red). $N=5000$, 
 $m=4$, $p_{\rm pa}=0.3$, $p_{\rm tr}=0.1$, $\tau=0.3$, $\gamma=1$, three graphs  and three SIR simulations on each graph.}
    \label{fig:discrete_SIR}
\end{figure}

We show the results for $\tau=0.3$, and $p_{\rm pa}=0.3, p_{\rm tr}=0.1$.
All other characteristics of the setup are the same as in Figure  \ref{fig:tau-estimate-pa}.
The right panel of Figure \ref{fig:discrete_SIR} shows 
how the number of \emph{SI edges per infected vertex} between households evolves over time, as well as the estimate 
of this quantity given by equation \ref{si_estimate}. From the figure we can see that the number of between-household SI edges cannot be estimated uniformly well 
as $C\cdot I_t S_t$ for some constant $C$, because the number of susceptible vertices decreases monotonically, while the 
number of SI edges per infected vertex first increases, and later on decreases.

\begin{figure}
 \centering
 \includegraphics[width=0.5\textwidth]{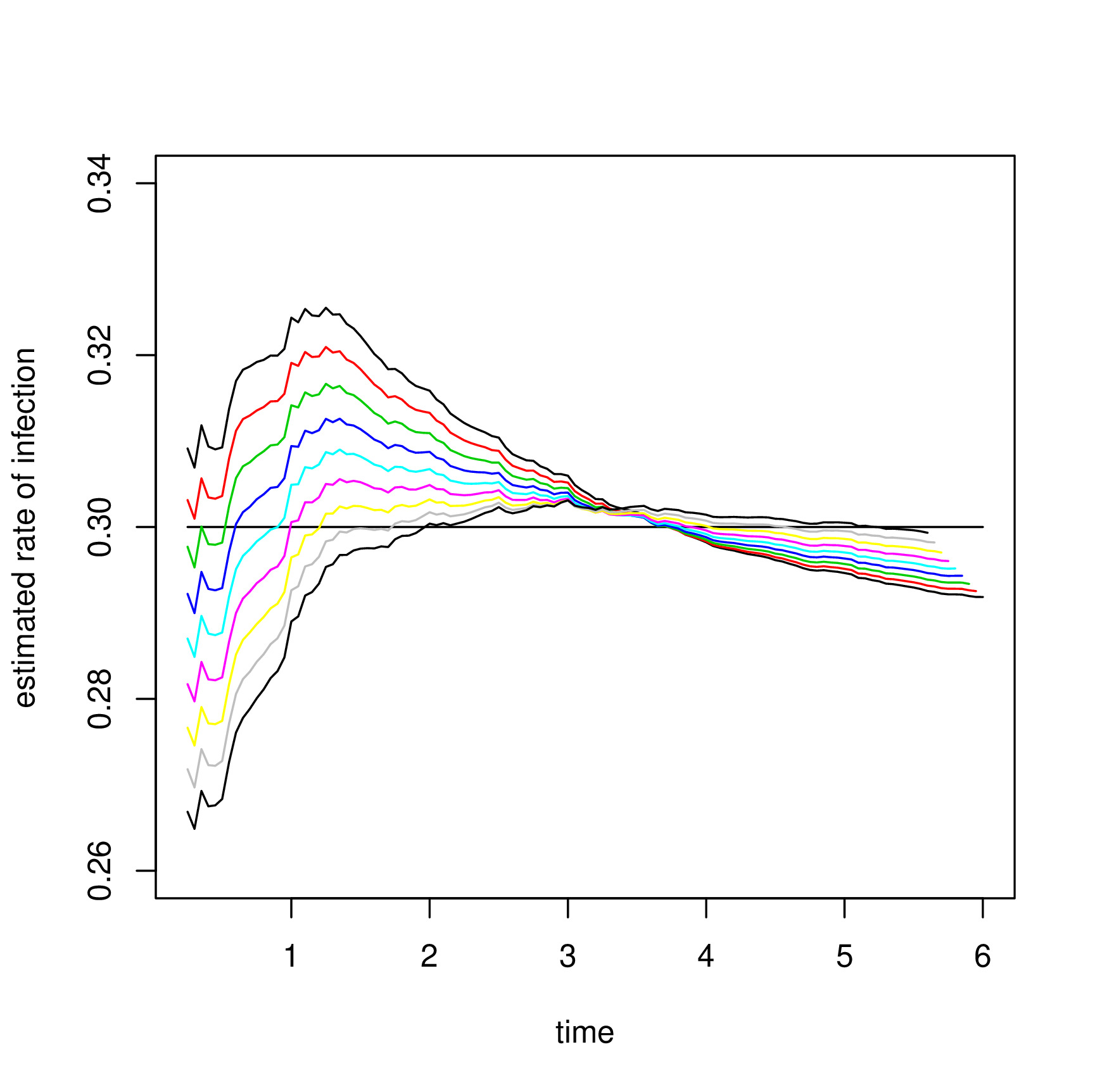}
 \caption{Estimated rate of infection, using the time-shifted estimate of equation
 \eqref{eq:shift} for the number of between-household SI edges. Top curve: no shift,
 and the shift time grows 
 on the left hand side from top to bottom, in steps of $0.05$. Each curve is the 
 average of nine simulations. $N=5000$,  $m=4$, $p_{\rm pa}=0.3$,
 $p_{\rm tr}=0.1$, $\tau=0.3$, $\gamma=1$.}
    \label{fig:shifted_tau}
\end{figure}

From the left panel of Figure \ref{fig:discrete_SIR} one can see that the peak of the
estimated number of between-household SI edges is lower than the true value, and also,
there seems to be a time-shift between the two curves. The idea arises to estimate the
number of SI edges between households at time $t$ by the formula
\begin{equation} \label{eq:shift}
\hat E^{SI,o}_t=I_{t+t_0}\cdot \bigg(d-\frac{wd}{wd+N_{\rm hh}-1}\bigg)\cdot \frac{S_{t+t_0}}{N},
\end{equation}
where $t_0$ is a suitably chosen shift. With this method, we need to
wait until time $t+t_0$ to estimate the contagion parameter $\tau$ at
time $t$, but if $t_0$ is small, this is not a severe limitation.
Figure \ref{fig:shifted_tau} shows the estimate of $\tau$ with 
various choices of $t_0$ ranging from $0$ to $0.4$ in steps of $0.05$. None of the curves seen in the figure 
perform well overall, but we may say that the shift seems to improve 
the quality of the estimate. However, the phenomenon would need more
careful exploration, and the choice of the shift size is another question.

\subsection{Parameter estimation with adjustment in the initial period of epidemics}
In this section we aim to improve the estimation of the $\tau$ parameter of the virus spread only on some initial phase of the whole process, by applying simulation based adjustment factors in $\hat E^{SI,o}_t$ the estimated numbers of SI edges outside households. Explicitly, we would like to find some 
$AdjFact(p_{u}, p_{tr}, p_{pa})$ adjustment factors, so that with the help of $\hat E^{SI,o}_t \times (1+ AdjFact(p_{u}, p_{tr}, p_{pa}) )$ we get better estimation.
The idea was motivated by the following considerations:
\begin{itemize}
    \item For real life applications, it is more realistic that one would like to estimate the infection rate right from the beginning of the pandemic, not only at the subsidence of the process. With the help of the $\hat E^{SI,o}_t$ estimation we introduced previously, $\hat \tau$ is able to match the real infection rate after a reasonably long period, however its use is suitable for earlier periods 
    only to some extent, which highly depends on the properties of the underlying random graph.
    \item Based on previous results on the convergence of $\hat\tau$   (Figure \ref{fig:tau-convergence-pa}), it can be concluded that generally in the initial period of the virus spread the number of SI edges are under-estimated, while later over-estimated, and the magnitude of error clearly depends on the structure of the random graph. If we aim to improve the estimation with the help of a single adjustment factor (which can be the function of parameters $p_u, p_{tr}, p_{pa}$, driving the graph structure), it is only possible on some fixed initial period of the virus spread, as this correcting factor would be highly time-dependent.
\end{itemize}

\begin{figure}
\begin{minipage}{0.5\textwidth}
    \centering
    \includegraphics[scale=0.4]{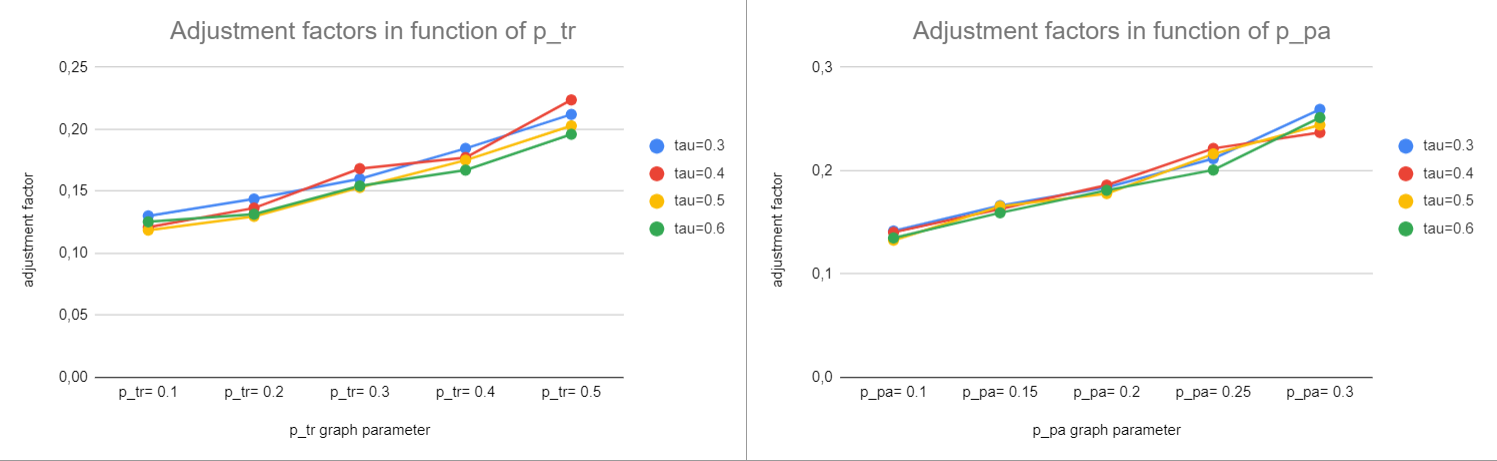}
    \end{minipage}
\caption{Calculated adjustment factors (mean of 100 independent trajectories) in function of $\tau$ and $p_{tr}$ or $p_{pa}$ parameters. As expected, lower $p_{u}$ values are yielding bigger estimation error.}
    \label{fig:grid}
\end{figure}

\begin{figure}
\begin{minipage}{0.5\textwidth}
    \centering
    \includegraphics[scale=0.25]{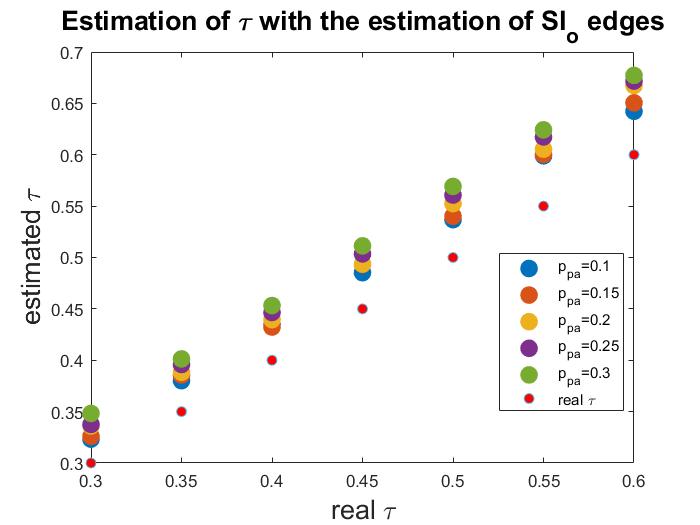}
    \end{minipage}
\begin{minipage}{0.5\textwidth}
    \centering
    \includegraphics[scale=0.215]{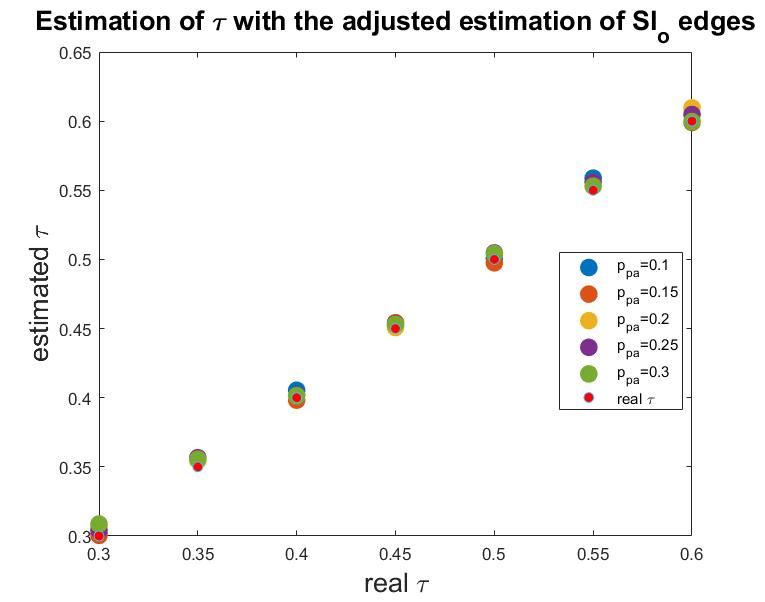}
    \end{minipage}
    
\caption{Estimate of the infection parameter $\tau$ for different polynomial random graphs  if the number of $SI$ edges between households is not known; with (right) and without adjustment (left). The graph has $N=5000$ vertices,  $m=4$,   $\gamma=1$; we run 50 SIR simulations, the dots are averages of the 50 estimates when $1/10$ of the epidemic events occured  ($p_{\rm tr}=0$ is fixed). With the adjusted approach, the estimated parameters were able to match real $\tau$ values.} 
    \label{fig:results_pa}
\end{figure}

Driven by the motivations above, firstly we tried to identify a function of adjustment, depending on parameters $t$, $p_u, p_{tr}$ and $p_{pa}$. 
As first of all, we wanted to grasp independently the pure effect of clustering coefficient and scale free property of the random graph, we 
changed $p_{tr}$ and $p_u$ 
and kept $p_{pa}=0$, while in the other parameter-dependent function we 
moved $p_{pa}$ and $p_u$ with $p_{tr}=0$. In both of the cases 
the following steps were carried out:

\begin{enumerate}
    \item For each $\tau$ and random graph parameter set, we tried to find the length of the initial period, on which it 
    is considered as meaningful to improve the estimation. On each trajectory we calculated the difference between estimated and real SI edges on the whole period, and also the fraction of non-estimated SI edges. Please refer to Figure \ref{fig:non-estimated-SI-1}. The length of this initial period was determined as the location of the biggest difference between real and estimated SI edges
    $len_{max}=\max_t \Big(\hat E^{SI,o}_t - E^{SI,o}_t\Big)$, as based on the simulations the fraction of non-estimated edges locally converges to some value. On each trajectory, we calculated the fraction of this time period regarding to the total epidemic events (infection or recovery), typically 0.1 times the time until the epidemic process stops. 
    
\begin{figure}
\begin{minipage}{0.5\textwidth}
    \centering
    \includegraphics[scale=0.15]{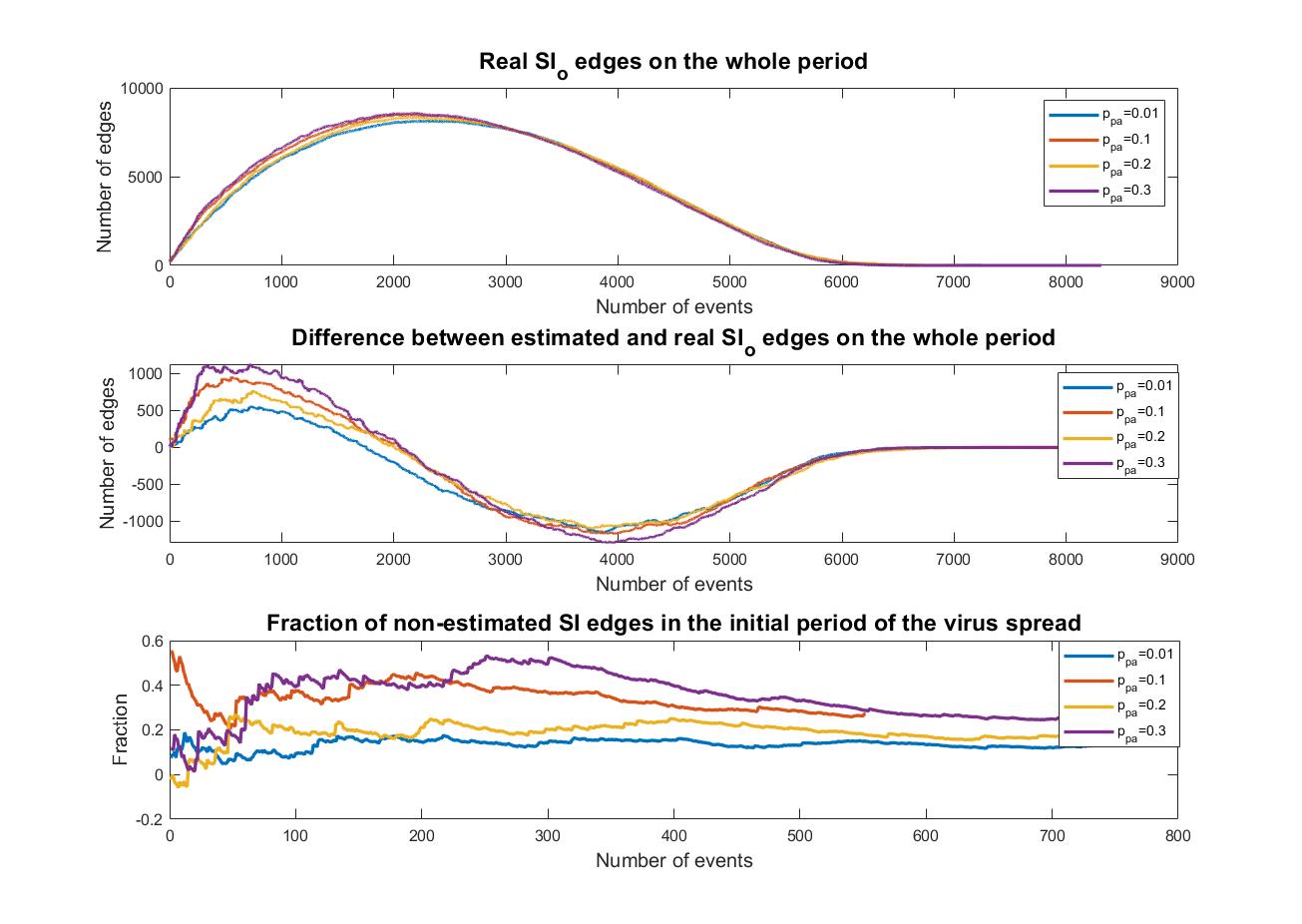}
    \end{minipage}
\begin{minipage}{0.5\textwidth}
    \centering
    \includegraphics[scale=0.15]{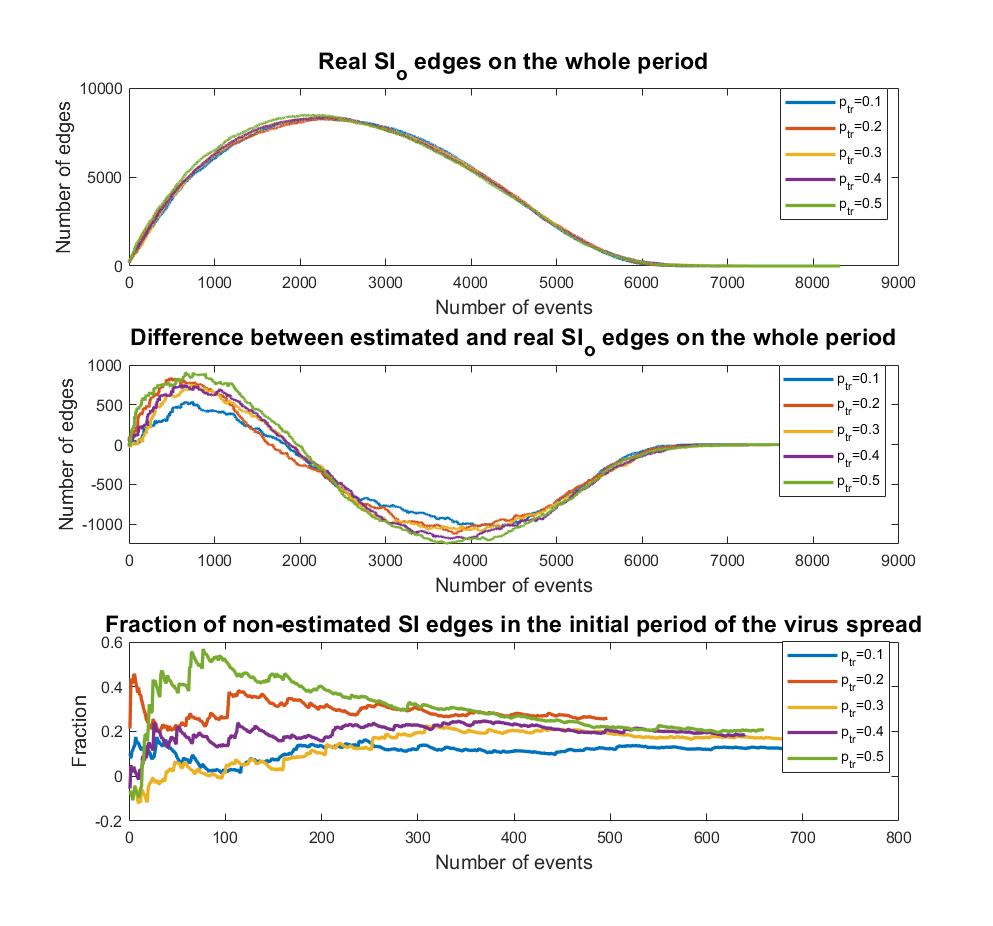}
    \end{minipage}
    \label{fig:non-estimated-SI-1}
    \caption{Error of $\hat E^{SI,o}_t$ estimation}
\end{figure}

\item On each testing trajectory we calculated the error of our initial estimation, which is the fraction of SI edges the estimation does not take into account $adj(t, p_{u}, p_{tr}, p_{pa})=\frac{\hat E^{SI,o}_t - E^{SI,o}_t}{\hat E^{SI,o}_t}$. Based on the definition of the estimation, it was logical to use the time-integral of $adj(t, p_{u}, p_{tr}, p_{pa})$ denoted by $AdjFact(p_{u}, p_{tr}, p_{pa})= \int_0^{len_{max}} adj(s, p_{u}, p_{tr}, p_{pa}) ds$.
\item For each different parameter set we generated 100 random graphs, simulated the virus spread process with some given $\tau$ and calculated the mean of the $AdjFact(p_{u}, p_{tr}, p_{pa})$ adjustment factors given on each trajectory. As a consequence created a parameter dependent adjustment factor, please refer to Figure \ref{fig:grid} for the results: It can be concluded that the calculated correcting factors do not depend on $\tau$, therefore the calculated values are meaningful for the improvement of $\tau$ estimation. Adjustment factors are monotonic both in function of $p_{pa}$ and $p_{tr}$ parameters. As expected, the preferential attachment component has very significant effect on the adjustment factors: e.g. with $p_u=0.7, p_{pa}=0.3, p_{tr}=0$ polynomial graph setup, the original estimation does not take into consideration the quarter of the real $SI_o$ edges, as the estimation is based on mean-field theory.
\item For the purpose of backtesting, we calculated the original estimation of the infection parameter $\tau$ 
 (similar to Figure \ref{fig:tau-estimate-pa}) based on data of the initial phase of the virus spread and contrasted it to the adjusted estimation. Please refer to Figure \ref{fig:results_pa} for the test results.
\end{enumerate}

\section{Deep learning of contagion dynamics}

\label{sec:gnn}

\subsection{The GNN framework}
The use of machine learning and in particular neural network-based approaches is becoming increasingly prominent in epidemiological modelling. Therefore, we compared the classical infection rate estimation that uses a maximum likelihood framework and which we discussed previously to methods relying on neural networks. Upon examining the literature, we found this particular type of parameter estimation an unadressed problem in case of graph-based epidemic models. However, the problem of epidemic inference rather than estimation of infection rate directly is somewhat more widely discussed \cite{biazzo2022bayesian} \cite{Chae2018} \cite{Menar2020}. One promising approach is a result by Murphy et al. \cite{gnn} who present a solution using a graph neural network (GNN). 

In this setup, the epidemic model is a discretized SIR process where each state transition event happens synchronously for every node at every timestep of given fixed length. Every contact can infect a neighbor independently with a fixed probabily, so the state transition at a given step only depends on the number of infected neighbors.

The GNN model can produce a prediction for the state of a node at a given timestep as a function of the previous state of the node itself and all of its neighbors. Therefore, the model uses precise information about the local structure of the underlying graph and the state of the epidemic locally (both within the graph and in time.) By averaging predictions, a probability distribution for state transition as a function of the number of infected neighbors $\ell$ is available, which the authors compared to the distribution obtained from maximum likelihood estimation. Since they make the ML estimation for each $\ell$ separately and thus also on smaller samples, they found the GNN-based method to produce a smoother fit to the ground truth.

The proposed goal in \cite{gnn} is to produce accurate predictions on transition probabilities in any network topology after training the GNN on just one particular topology. Perhaps not surprisingly, the authors found scale-free Barabási-Albert (BA) graphs to provide the best training dataset as it is considerably heterogenous in terms of local structure.

For this setup the GNN architecture is the following:
\begin{itemize}
\item Input layers (Linear(1,32), ReLU, Linear(32, 32), ReLU): these transform the node states and possible node attributes into vectors.
\item Attention layers (in this case 2 of them in parallel): these aggregate the node variables with the features of its neighbors, and the possible edge attributes. 
\item Output layers (Linear(32, 32), ReLU, Linear(32, 2) Softmax): compute the outcomes of the nodes.
\end{itemize}
This architecture has altogether $6698$ trainable parameters, and several hyperparameters to be optimized which include time series length,
network size, resampling time, and importance sampling bias. 

\subsection{Deep learning of dynamics for clustered graphs}
In this section we present our results regarding the performance of the neural network based approach presented by Murphy et al. \cite{gnn}, as a function of average clustering of the network in the epidemic model. As stated before, in the stochastic epidemic model the infection dynamics is equivalent to a probability distribution of transitioning from a susceptible to infected state, as a function of the number of infected neighbours (thus it is a function of local structure).

The neural network is trained with generated data from an epidemic model with a given network topology. We extended the setup to include highly clustered networks (i.e. with asymptotically non-vanishing average clustering coefficient) in both GNN training and testing datasets. The model presented in Holme et al. 2001 \cite{clustered-ba} possess both scale-free structure and high average clustering. 
Although this model is slightly different from the one we used in the previous sections, we chose to use it out of convenience, as \textit{Python} module \href{https://networkx.org/}{\textit{NetworkX}} already contained an algorithm generating graphs based on said model. Here, the graph-generating algorithm 
includes an additional triangle-forming step after the preferential attachment step for the addition of each new node. The model produces a wide range of average clustering 
as a function of a tunable parameter -- the triangle forming probability -- whilst preserving the scale-free property of the Barabási-Albert model. To fit it to our model used in the MLE-based approach we added a further layer of households to the network consisting of disjoint cliques of a given size ($N_{\rm hh}=5$ as in the previous sections).\\
We compared the performance of the neural network in cases of training datasets from network models with different average clustering. The performance was evaluated by calculating the $l_{2}$-error of the GNN prediction on contagion dynamics compared to the ground truth provided by the simulated datasets.\\

\begin{figure}
    \centering
    \includegraphics[width=0.7\linewidth]{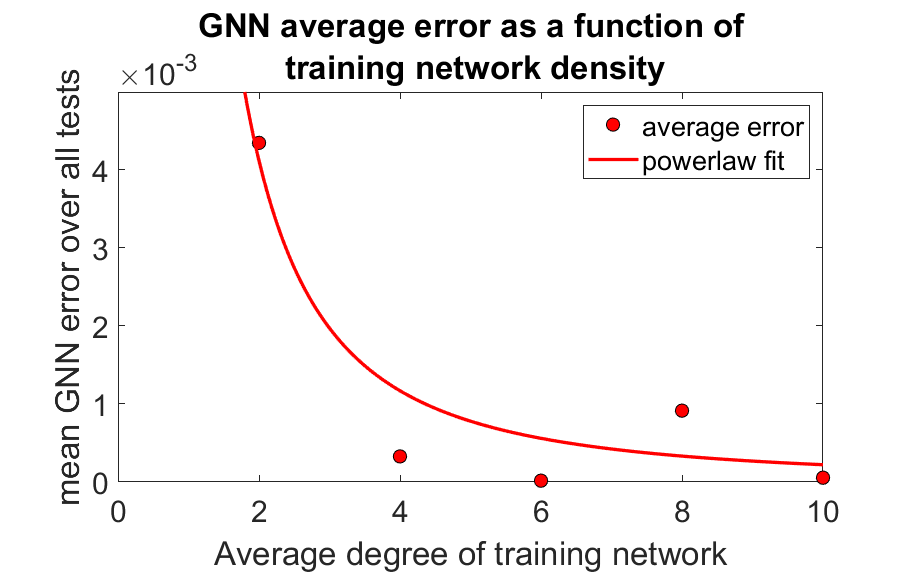}
    \caption{Using BA networks for GNN training, a clear relationship between training network density and GNN performance can be observed.}
    \label{fig:degree_dep}
\end{figure}

\begin{figure}%[ht]g
    \centering
    \includegraphics[width=0.7\linewidth]{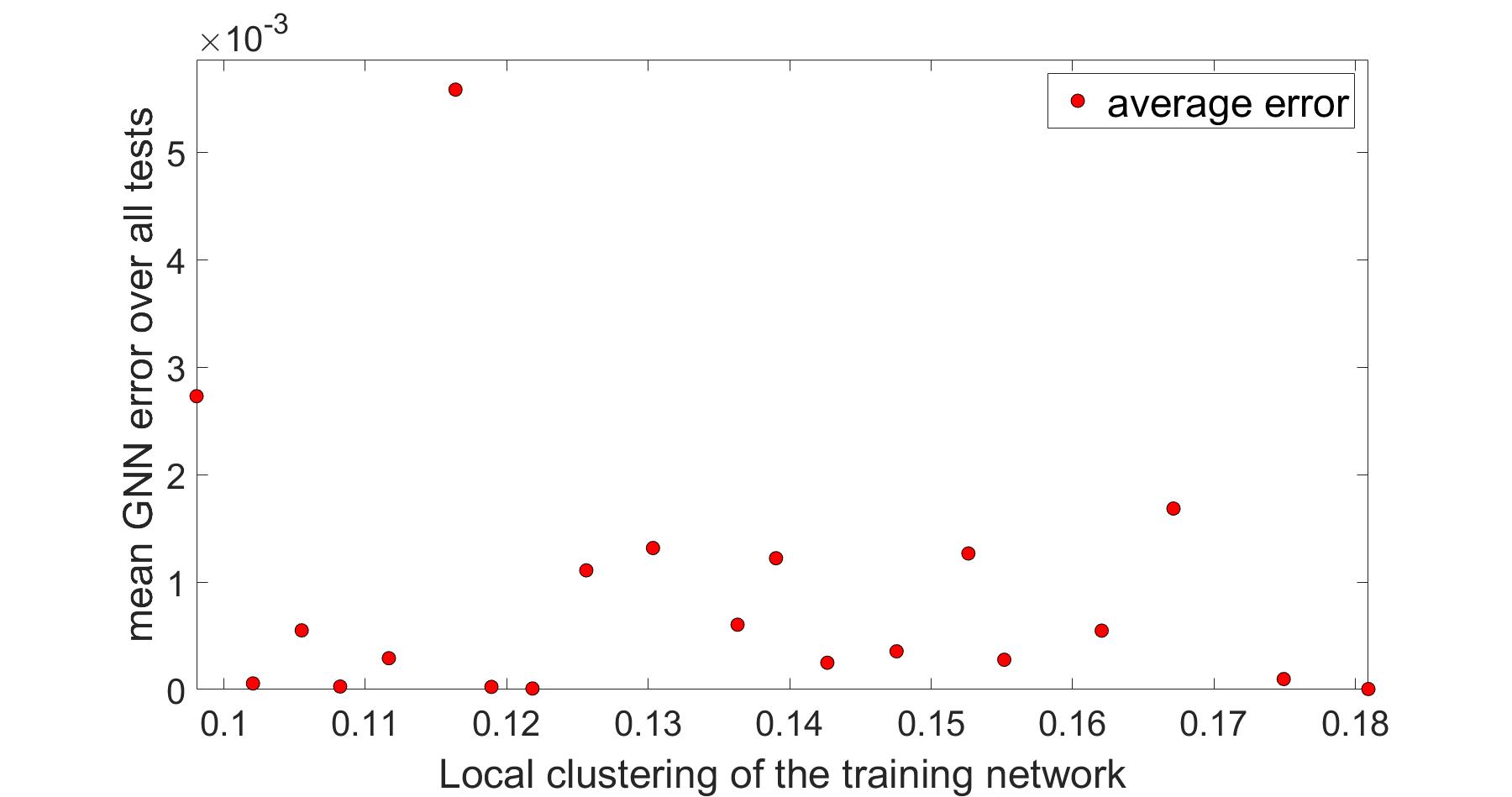}
    \caption{GNN performance as a function of average clustering of the graph generating the training dataset. The average degree of the graphs 
    was fixed ($\overline{d} = 13.74$, $\sigma = 0.55$).}
    \label{fig:fixed_degree}
\end{figure}

\begin{figure}
\begin{minipage}{0.5\textwidth}
\centering
  \includegraphics[scale=0.1]{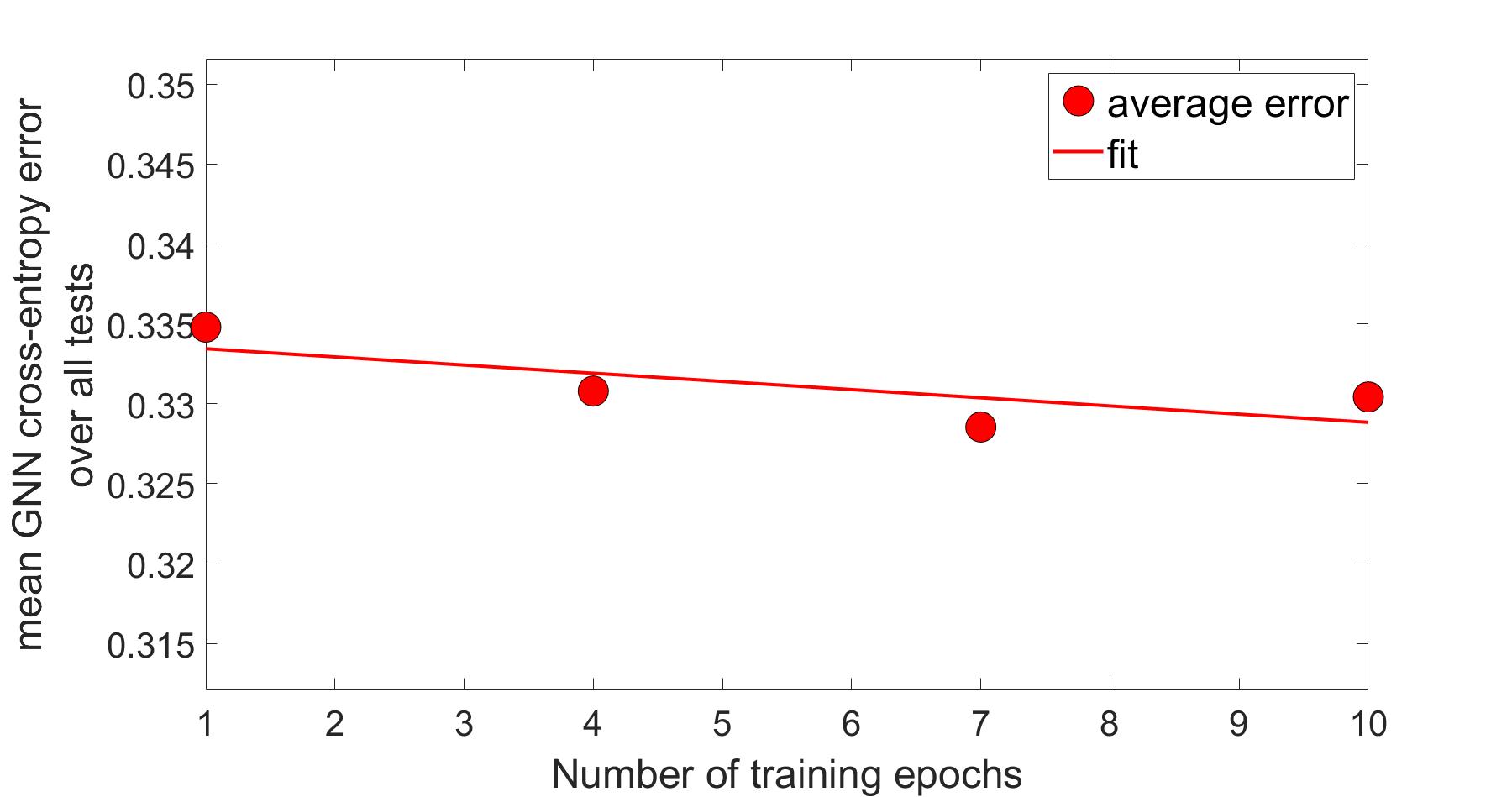}  
  \end{minipage}
\begin{minipage}{0.5\textwidth}
\centering
  \includegraphics[scale=0.1]{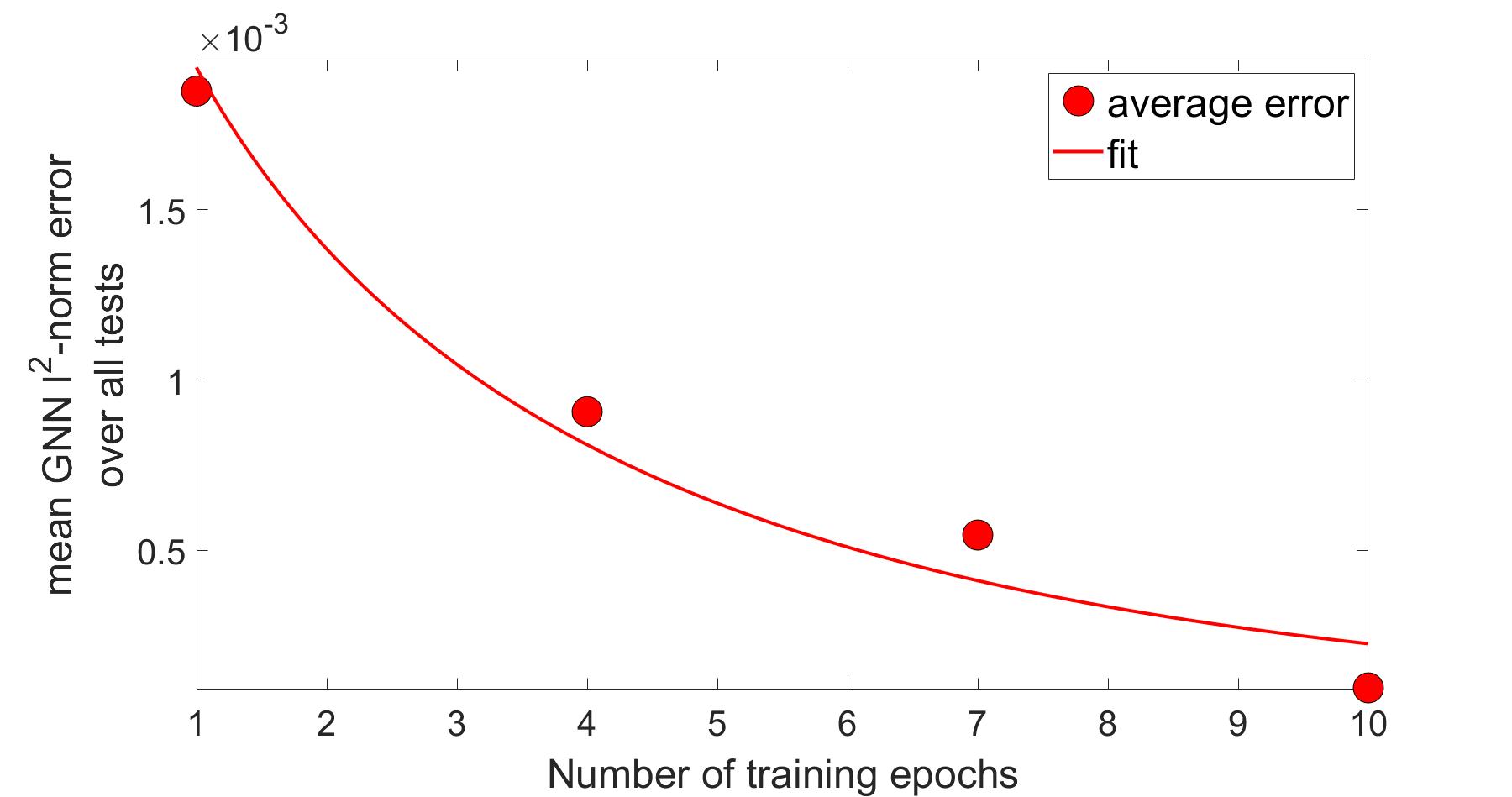}  
  \end{minipage}
  \caption{Evaluating GNN performance as a function of the number of training epochs, using the cross-entropy value described in \cite{gnn} (left), and using $l_{2}$ norm (right). The slope on the left is $-1.873\cdot10^{-4}$ and the fitted curve is $y = f(x) = c\cdot e ^{\alpha x^{\beta}}$, with $c = 0.0032$, $\alpha = -0.5008$ and $\beta = 0.7219$ minimizing the mean-squared error of the functional form $f$.}
  \label{fig:cross_entropy}
  \end{figure}

\begin{figure}
  \centering
   \includegraphics[width=0.6\linewidth]{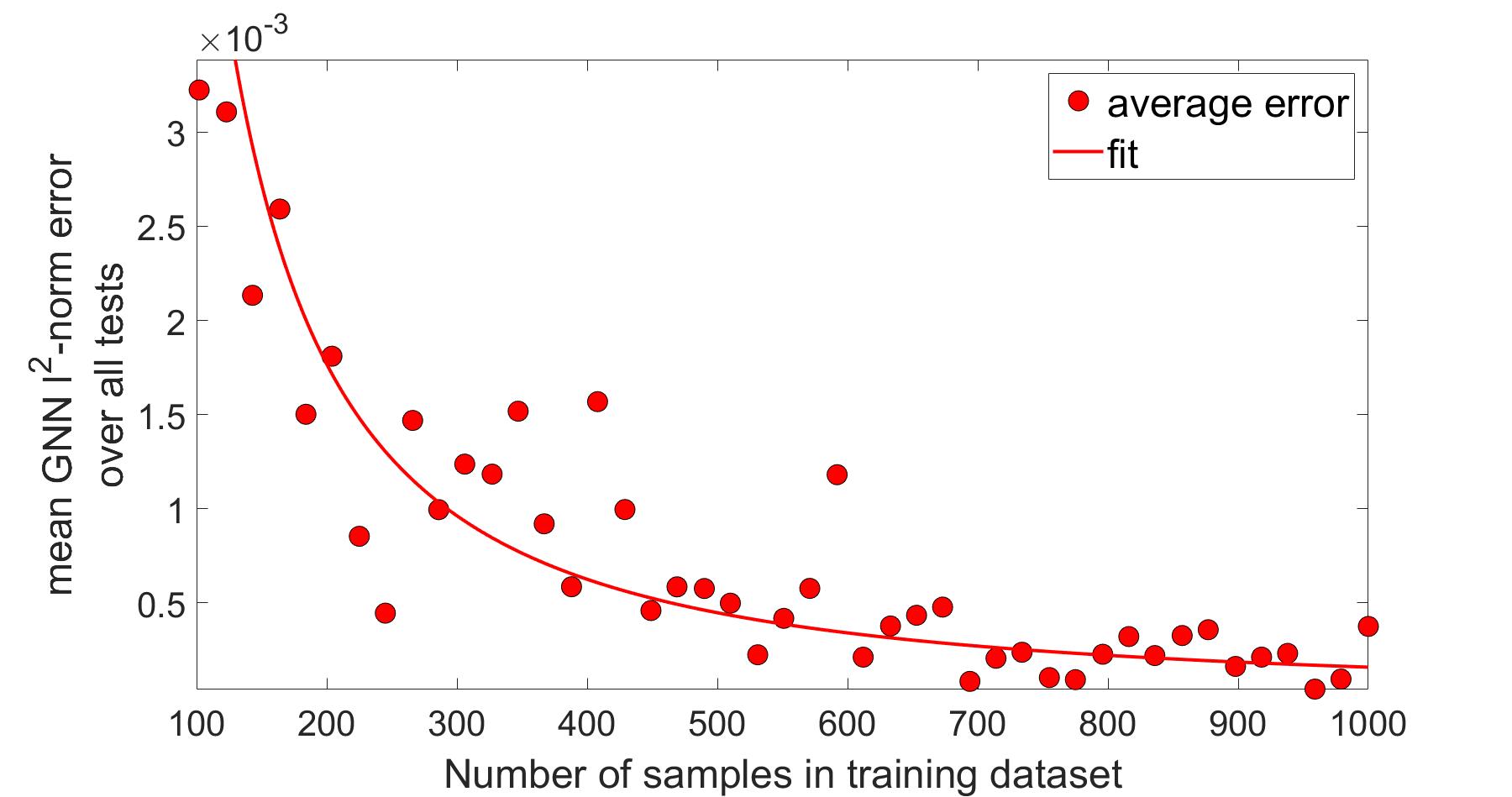}
    \caption{Evaluating GNN performance as a function of the number of datapoints in the training dataset. Featured is the curve $y = f(x) =c\cdot x^{\beta}$, with $c = 2.4068\cdot10^{4}$ and $\beta = -3.6853$ as a function, minimizing mean-squared error of the $f$ functional form.} 
    \label{fig:num_sample}
\end{figure}

When using data generated from networks from any model class for GNN training we found a clear relationship between network density and GNN performance as seen in Figure \ref{fig:degree_dep}.
Dense networks offered better training datasets in a given parameter range. Controlling for this parameter, the previously observed relationship between clustering and performance proved to be unclear as seen in Figure \ref{fig:fixed_degree}. \\
Evaluating the training using the evaluation function provided by \textit{PyTorch}, and the cross-entropy value described in \cite{gnn} (i.e. 
$$\mathcal{L}(\mathbf{\Theta})=\sum\limits_{t \in \mathcal{T}'}\sum\limits_{v_{i} \in \mathcal{V}'(t)} \frac{w_{i}(t)}{Z'}L\left(y_{i}(t),\hat{y}_{i}(t)\right),$$ where $\mathbf{\Theta}$ are the GNN parameters, $\mathcal{T}'$ and $\mathcal{V}'(t)$ are the training time set and node set, $w_{i}(t)$ are importance weights, $Z'$ is a normalization factor, $y_{i}(t)$ and $\hat{y}_{i}(t)$ are the real transition probabilities  to the possible states and the GNN output respectively, while $L(y_{i},\hat{y}_{i})=-\sum\limits_{m}y_{i,m}\log\hat{y}_{i,m}$ is the local cross-entropy loss function with $y_{i,m}$ the $m$th element of the output vector) we can observe non-monotonic and nearly constant losses at any stage of the GNN training, while the $l_{2}$ error decreased monotonically as seen in Figure \ref{fig:cross_entropy}, i.e.. This was the motivation for using the $l_{2}$ metric as a reliable means of evaluation. \\
We estimated the dependence of network performance on the number of datapoints in the neural network training datasets as a powerlaw with exponent $\beta = -3.6853$ and constant $c = 2.4068\cdot10^{4}$ as seen in Figure \ref{fig:num_sample}.

\section*{Conclusions} 

In this paper we considered a two-layer random graph, consisting of  a fixed, deterministic household structure and a preferential attachment model based on \cite{clustering}, added independently. The parameters of the second layer had a significant effect on the degree distribution and the average local clustering coefficient. Given these random graphs with different structural properties, we proposed two methods to estimate the infection parameter in an SIR model.  The first, simpler method does not use detailed information about the graph (only quantities like  the average degree of a vertex outside its household or the actual number of infected individuals), while the second method is based on a graph neural network, and uses much more detailed information about the graph (e.g. information about the neighbors). Hence, in general, it is not easy to compare the two methods, as they are elaborated for significantly different setups. However, in both cases we may ask whether the structure of the graph (degree distribution, clustering coefficients) has a visible effect on the quality of the estimates. 

In the first case we can conclude that 
increasing the weight of the preferential attachment component, which is closely related to the proportion of vertices with very large degree, leads to larger error in our estimates. As for the GNN method, we observed that the edge density had a significant effect, graphs with larger edge density formed more optimal training sets. On the other hand, 
the local clustering coefficient does not seem to have a significant effect on the errors of the estimates in either case. 

Hence we may conclude that understanding the effect of degree distribution and edge density might be more important than other structural properties, such as the clustering coefficient. On the other hand, it is an open question what kind of neural networks or machine learning methods are the most efficient if we have less information about the graph and the epidemic spread process, and which statistics are indeed necessary to obtain good estimates.

\section*{Acknowledgement}
This research was supported by the National Research, Development and Innovation Office within the framework of the Thematic Excellence Program 2021 - National Research Sub programme: “Artificial intelligence, large networks, data security: mathematical foundation and applications”.

\section*{Declarations}

\begin{itemize}
\item {\bf Funding.} This research has been implemented with the support provided by the Ministry of Innovation and Technology of Hungary from the National Research, Development and Innovation Fund, financed under the  ELTE TKP 2021-NKTA-62 funding scheme.
\item {\bf Author's contribution.} \'Agnes Backhausz 
worked on the computer simulations mainly about parameter sensitivity, and 
on the text and figures in Sections 1-4. Edit Bogn\'ar wrote and run the computer simulations which lead to the adjusted estimates in Section 4, and worked on the text and figures of this section. Vill\H o Csisz\'ar wrote and run the computer simulations about parameter sensitivity and the time-shifted estimate, and worked on the text and figures for Sections 2-5. Damj\'an T\'ark\'anyi worked on the computer simulations and the programme codes necessary for the  application of the graph neural network for this problem, and hence prepared and elaborated the text and figures in Section 5. Andr\'as Zempl\'eni worked on the overall concept of the paper, on choosing the applied methods, and on the formulation of the text and the figures. 
\item {\bf Availability of Data and Materials.} Data availability is not applicable. The programme codes are available by contacting the corresponding author (\'Agnes Backhausz). 
\item {\bf Competing interests.} The authors declare that they have no competing interests.
\end{itemize}

\bibliographystyle{plain}
\bibliography{elsocikkv2}% common bib file

\end{document}